\documentclass[11pt,a4paper]{iopart}
\usepackage[utf8]{inputenc}
\usepackage{url}
\usepackage{graphicx}
\usepackage{iopams}
\usepackage{setstack}
\usepackage{color}
\usepackage{bm}

\makeatletter
\long\def\dddddot#1{%
  {\mathop {#1}\limits ^{\vbox to-1.4\ex@ {\kern -\tw@ \ex@ \hbox {\normalfont .....}\vss }}}%
}
\long\def\multidots#1#2{%
  \count@=0
  {{\mathop {#2}\limits ^{\vbox to-1.4\ex@ {\kern -\tw@ \ex@ \hbox {\normalfont %
  \loop%
  \ifnum#1>\count@%
  .%
  \advance\count@ by1%
  \repeat%
  }\vss }}}}%
}
\makeatother

\newcommand{\udt}[3]{#1^{#2}_{\phantom{#2}#3}}
\newcommand{\udut}[4]{#1^{#2\phantom{#3}#4}_{\phantom{#2}#3\phantom{#4}}}

\newcommand{\dut}[3]{#1_{#2}^{\phantom{#2}#3}}
\newcommand{\dudt}[4]{#1_{#2\phantom{#3}#4}^{\phantom{#2}#3}}

\begin{document}

\title[Cosmological viable models in $f(T,B)$ theory as solutions to the $H_0$ tension]{Cosmological viable models in $f(T,B)$ theory as solutions to the $H_0$ tension}

\author{Celia Escamilla-Rivera}
\address{Instituto de Ciencias Nucleares, Universidad Nacional Aut\'{o}noma de M\'{e}xico, 
Circuito Exterior C.U., A.P. 70-543, M\'exico D.F. 04510, M\'{e}xico.}
\ead{celia.escamilla@nucleares.unam.mx}

\author{Jackson Levi Said}
\address{Institute of Space Sciences and Astronomy, University of Malta, Msida, MSD 2080, Malta.}
\address{Department of Physics, University of Malta, Msida, MSD 2080, Malta.}
\ead{jackson.said@um.edu.mt}

\date{\today}

\begin{abstract}
In this work we present a further investigation about Teleparallel Gravity Cosmology. We demonstrate that according to the current astrophysical data (CC+Pantheon+BAO samplers with late universe measurements SH0ES+H0LiCOW), a $f(T,B)$ theory can provide another interpretation to the oscillatory behaviour of the dark energy equation of state when applied to late times. The four $f(T,B)$ cosmological viable models proposed here can undergo an epoch of late-time acceleration and
reproduce quintessence and phantom regimes with a transition along the phantom-divided line, making this theory a good approach to modify the standard $\Lambda$CDM model.
\end{abstract}

\maketitle


\section{Introduction}
It is well know that the $\Lambda$CDM cosmological model is motivated by astounding successes in describing the Universe at all scales where observations can be made \cite{misner1973gravitation,Clifton:2011jh,Ishak:2018his}. Through the proposal of cold dark matter, this cosmological model can adequately describe the dynamics of galaxies, and through the effects of dark energy, the cosmic acceleration of the Universe \cite{Baudis:2016qwx,Bertone:2004pz}. However, despite great efforts, dark matter remains undetected and the cosmological constant description via dark energy continues to have numerous problems associated with it \cite{RevModPhys.61.1}.

On one hand, recently, the effectiveness of the $\Lambda$CDM model in explaining precision cosmology observations has been called into question. This is primarily through the so-called $H_0$ tension problem which is a discrepancy between the predicted value of $H_0$ from the early Universe and its observed value from the local measurements. First reported as a serious tension by the Planck collaboration in \cite{Aghanim:2018eyx,Ade:2015xua}, the tension has since grown by means of strong lensing measurements from the H0LiCOW ($H_0$ lenses in Cosmograil's wellspring) collaboration \cite{Wong:2019kwg} and from Cepheids via SH0ES (Supernovae $H_0$ for equation of state) \cite{2006hst..prop10802R}. Meanwhile, measurements based on the tip of the red giant branch (TRGB, Carnegie-Chicago Hubble Program) have yielded a lower $H_0$ tension \cite{Freedman:2019jwv}. There also exist novel methods of determining the value of $H_0$ through gravitational wave astronomy \cite{Graef:2018fzu,Abbott:2017xzu} that may shed light on the problem in the near future.

On the other hand, by construction, $\Lambda$CDM is based on taking Einstein's theory of General Relativity (GR) and modifying the matter content part of the theory to satisfy observational demands. However, it may also be the case that the gravity content of these cosmological model needs to be corrected to account for current observationals. There have been a plethora of theories that have been proposed which have had varying success in explaining the Universe. However, by and large these theories mainly rely on considering gravity through the prism of the Levi-Civita connection which expresses gravitation through the curvature of spacetime, as in GR. In this work, we consider Teleparallel Gravity (TP) which differs from GR in that it manifests gravity through torsion rather than curvature \cite{Weitzenbock1923}, which is achieved by replacing the Levi-Civita connection with the Weitzenb\"{o}ck connection.

At the level of the gravitational action, GR and TG can be made to be equal up to a boundary term, this is the so-called \textit{Teleparallel equivalent of General Relativity} (TEGR). Straightforwardly, TEGR will produce the same field equations as GR, but the ensuing modifications that can be constructed will naturally be distinct. TG also has a number of other advantageous features such as its similarity to Yang-mills theory \cite{Aldrovandi:2013wha} giving it an added particle physics dimension. It is also possible to define a gravitational energy-momentum tensor in TG \cite{Blixt:2018znp,Blixt:2019mkt} which means separating inertia and gravitation. However, this remains an open question in teleparallel gravity more generally \cite{deAndrade:2000kr,Maluf:1996kx,Maluf:2012na,Bamba:2015dla,Ulhoa:2010wv} and is an interesting topic for further development. TG is more regular than GR in that it does not require the introduction of a Gibbons--Hawking--York boundary term in order to produce a well-defined Hamiltonian formulation \cite{Cai:2015emx,Maluf:2013gaa,Ferraro:2018tpu}. Even more, TG is an interesting theory of gravity since it does not necessarily require the equivalence principle to hold, that is, unlike GR where this is a fixed feature, TG would survive a violation of this principle \cite{Aldrovandi:2004fy}, if future observations were to reveal such a violation.

As with GR, TG can be readily modified in numerous routes: the TEGR Lagrangian is the so-called torsion scalar $T$ (discussed in \S.~\ref{sec:TG}), where we can immediately generalised to $f(T)$ theory which has generally second-order field equations unlike $f(R)$ gravity \cite{RevModPhys.82.451,Faraoni:2008mf,Capozziello:2011et} that is a fourth-order theory. Analogously, many other approaches that appear in theories based on the Levi-Civita connection can analogously be constructed in TG. For example, the Gauss-Bonnet scalar, $G$, can be equivalently constructed in TG, as $T_G$, and used to construct other modified theories of gravity \cite{Kofinas:2014owa,Bahamonde:2016kba,Capozziello:2016eaz,Zubair:2015yma,Jawad2015a}. In this last example, we would be comparing $f(R,G)$ \cite{Odintsov:2018nch,Atazadeh:2013cz} gravity to $f(T,T_G)$ gravity \cite{Kofinas:2014aka,Kofinas:2014daa,delaCruz-Dombriz:2017lvj,delaCruz-Dombriz:2018nvt,Chattopadhyay:2014xaa,Waheed:2015bga}. Here, the ensuing equations of motion would be fourth-order due to appearance of the Gauss-Bonnet scalar invariant. 

In theories constructed out of the Levi-Civita connection, the Lovelock theorem \cite{Lovelock:1971yv} heavily limits to possibilities of forming second-order theories of gravity. However, given the ease with which second-order theories are formed in TG makes it an ideal platform on which to form such theories. In many cases, no curvature-based corresponding model can be formed. One example of this is new general relativity (NGR) which takes the irreducible components of the torsion tensor and forms a general theory in which these irreducible components form GR for specific linear coefficients only \cite{Hayashi:1979qx,Pereira:2001xf,Blixt:2018znp}. For other choices of combinations of such irreducible contributions, no GR analog exists. 

Another interesting property that relates GR and its teleparallel equivalent, TEGR, is where the Ricci scalar is equivalent to the torsion scalar added with a total divergence term $B$ (boundary term) (discussed in \S.~\ref{sec:TG}). In this way, an interesting model can be formed where these scalars contribute arbitrarily. This is $\tilde{f}(T,B)$ gravity where the torsion scalar and boundary term contribute independent of each other through the arbitrary function, $\tilde{f}$ \cite{Bahamonde:2015zma,Capozziello:2018qcp,Bahamonde:2016grb,Paliathanasis:2017flf,Farrugia:2018gyz,Bahamonde:2016cul,Bahamonde:2016cul,Wright:2016ayu}. Clearly, since these scalars form the Ricci scalar, this model will limit to $f(R)$ for particular choices of the model. However, in general this will form a much richer theory on which to form cosmological models. In terms of derivatives, the torsion scalar and boundary term embody the second- and fourth-order contributions to the field equations. This is another way of viewing the fourth-order metric derivative $f(R)$ equations of motion.

TG is different to curvature-based models of gravity in that its dynamical field is constructed on the tetrad formulation of gravity \cite{misner1973gravitation} which incorporates the appearance of an inertial spin connection. For a flat FLRW cosmology, we can make a choice of tetrad such that the spin connection is allowed to be zero. Thus, we then construct the Friedmann equations which are straightforwardly identity to GR in their TEGR setting \cite{Pereira:2013qza}. Our interest lies in taking the $\tilde{f}(T,B)$ extension to TEGR \cite{Bahamonde:2017bps}, which in general forms fourth-order field equations in terms of tetrad derivatives. In this work, we consider four literature models of $\tilde{f}(T,B)$ gravity in order to study their cosmological dynamics at late-times. In this way, we attempt to mimick the behaviour of dark energy and reduce the $H_0$ tension. This problem has been well-studied in generalized TG such as the works Refs.\cite{Yan:2019gbw,Nunes:2018xbm}

Throughout this work, Latin indices are used to refer to coordinates on the tangent space, while Greek indices refer to general manifold coordinates. Also, the metric signature is $\eta_{\mu\nu} = \mbox{diag}(-1, 1, 1, 1)$. The outline of the paper is as follows: in \S.~\ref{sec:TG} we describe the TG background theory in order to set the equivalence with GR. In \S.~\ref{sec:f(T,B)_cosmo} we explore a flat homogeneous and isotropic cosmology in the $\tilde{f}(T,B)$ gravity setting in order to derive a generic equation of state for the theory. In \S.~\ref{sec:data_analysis}, we describe the observational data being used to constrain the cosmological parameters for the models being investigated. \S.~\ref{sec:models} contains the analyses of the $\tilde{f}(T,B)$ models being proposed. The observational constraints will be obtained using astrophysical data as galaxy ages sampler, BAO samples and supernovae Type Ia current sampler (Pantheon). The statistical results are shown for each model. Finally, a summary and conclusion of our work is given in \S.~\ref{sec:conclusions}. An Appendix with all the general calculations for each proposed cosmological model can be found at the end.

\section{Teleparallel Gravity background}
\label{sec:TG}

General Relativity (GR) expresses gravitation in terms of curvature by means of the Levi-Civita connection, $\mathring{\Gamma}{}^{\sigma}{}_{\mu\nu}$ \cite{misner1973gravitation}. However, Riemann geometry contains other means of geometric deformation which can be used to describe gravity. In fact, there exists a trinity of characterizations of gravity such that GR can be reproduced at the level of the field equations in a particular limit \cite{BeltranJimenez:2019tjy}. In this work, we consider the setting of TG \cite{aldrovandi1995introduction,Cai:2015emx,Krssak:2018ywd}. TG is fundamentally distinct from curvature-based descriptions of gravity in that the Levi-Civita connection is replaced with the Weitzenb\"{o}ck connection, $\Gamma^{\sigma}{}_{\mu\nu}$, which is a curvatureless connection that observes the metricity condition \cite{Weitzenbock1923,Aldrovandi:2013wha}. The Weitzenb\"{o}ck connection is defined by
\begin{equation}\label{eq:weitzenbockdef}
\udt{\Gamma}{\sigma}{\mu\nu} := \dut{e}{a}{\sigma}\partial_\mu \udt{e}{a}{\nu} + \dut{e}{a}{\sigma}\udt{\omega}{a}{b\mu}\udt{e}{b}{\nu}\,,
\end{equation}
where $\udt{e}{a}{\rho}$ is the tetrad field ($\dut{e}{a}{\mu}$ being the transpose), and $\udt{\omega}{a}{b\mu}$ the spin connection. This is the most general linear affine connection that is both curvatureless and satisfies the metricity condition \cite{aldrovandi1995introduction}. The tetrad relates the general manifold and the tangent (inertial) space, represented by the inertial Latin indices and the general manifold Greek indices. On the other hand, the spin connection appears in order to conserve the invariance of teleparallel theories under Local Lorentz Transformations (LLT) \cite{Li:2010cg}. Due to their inertial nature, the effect of the spin connection components can be understood as being entirely local \cite{Maluf:2013gaa,Maluf:2015cna,Krssak:2015rqa}. Thus, the spin connection incorporates the LLT freedom for any choice of theory based on the Weitzenb\"{o}ck connection. The spin connection is thus flat and can even be set to zero for a particular choice of Lorentz frame, which will be related by Lorentz matrices in all other frames \cite{RevModPhys.48.393}. GR also has an associated spin connection, but this is mainly hidden in the inertial structure of the theory \cite{aldrovandi1995introduction}. Together, the tetrad and its associated spin connection play the same role as the metric tensor in curvature-based theories of gravity.

The spin connection can be determined by considering the the full breath of LLTs (Lorentz boosts and rotations), where the tetrad is transformed by its inertial index through
\begin{equation}
\udt{e'}{a}{\mu}=\udt{\Lambda}{a}{b}\udt{e}{b}{\mu}\,,
\end{equation}
where $\udt{\Lambda}{a}{b}$ is a LLT. The spin connection can then be represented as the combination of completely inertial LLTs in the form \cite{Krssak:2015oua}
\begin{equation}
\udt{\omega}{a}{b\mu} = \udt{\Lambda}{a}{c}\partial_{\mu}\dut{\Lambda}{b}{c}\,,
\end{equation}
which preserves the LLT invariance of the theory as a whole. However, there also exist so-called good tetrad choices which produce vanishing spin connection components \cite{Tamanini:2012hg,Bahamonde:2017wwk}. Given the invariance of the theory under LLTs, all consistent tetrad and spin connection choices will be dynamically equivalent.

Thus, the metric tensor, $g_{\mu\nu}$, expresses geometric deformation through distance measurements, while the tetrad, $\udt{e}{a}{\mu}$ relates the tangent space with the general manifold. For consistency, they also observe the relations \cite{Aldrovandi:2013wha}
\begin{eqnarray}
\udt{e}{a}{\mu}\dut{e}{b}{\mu} = \delta^a_b\,,&\quad
&\udt{e}{a}{\mu}\dut{e}{a}{\nu} = \delta^{\nu}_{\mu}\,,
\end{eqnarray}
which form the orthogonality conditions of the tetrad fields. Naturally, the tetrad fields can be used to transform between the inertial Minkowski metric and a general manifold through
\begin{eqnarray} \label{Min_trans_Gen}
g_{\mu\nu} = \udt{e}{a}{\mu}\udt{e}{b}{\nu}\eta_{ab}\,,&\quad
&\eta_{ab} = \dut{e}{a}{\mu}\dut{e}{b}{\nu}g_{\mu\nu}\,,
\end{eqnarray}
where the tetrad can be seen to replace the metric tensor as the fundamental dynamical object of the theory. The position dependence of these relations is being suppressed for brevity's sake.
One important point is that the curvature measured by the Riemann tensor will always vanish in TG because the Weitzenb\"{o}ck connection is curvatureless, while the torsion will depend on the specific form of the tetrad and its associated spin connection. In this scenario, torsion is represented by the anti-symmetric part \cite{Bahamonde:2017wwk}
\begin{equation}
\udt{T}{\sigma}{\mu\nu} :=  2\udt{\Gamma}{\sigma}{[\mu\nu]}\,,
\end{equation}
which is a measure of the field strength of gravitation, and where square brackets represent the anti-symmetric operator, $A_{[\mu\nu]}=\frac{1}{2}\left(A_{\mu\nu}-A_{\nu\mu}\right)$. $\udt{T}{\sigma}{\mu\nu}$ is called the torsion tensor and transforms covariantly under both diffeomorphisms and LLTs.
Analogous to the Riemann tensor, the torsion tensor is a measure of torsion for a gravitational field. However, other useful tensors can also be defined, such as the contorsion tensor which is the difference between the Levi-Civita and Weitzenb\"{o}ck connections \cite{Cai:2015emx,RevModPhys.48.393}
\begin{equation}
\udt{K}{\sigma}{\mu\nu} := \Gamma^{\sigma}{}_{\mu\nu} - \mathring{\Gamma}{}^{\sigma}{}_{\mu\nu} = \frac{1}{2}\left(\dudt{T}{\mu}{\sigma}{\nu}+\dudt{T}{\nu}{\sigma}{\mu}-\udt{T}{\sigma}{\mu\nu}\right)\,.
\end{equation}
Naturally, this plays a crucial role in relating TG with Levi-Civita based theories. Another important ingredient in forming a teleparallel theory of gravity is the so-called superpotential which is defined as
\begin{equation}
\dut{S}{a}{\mu\nu} := \frac{1}{2}\left(\udt{K}{\mu\nu}{a} - \dut{e}{a}{\nu}\udt{T}{\alpha\mu}{\alpha} + \dut{e}{a}{\mu}\udt{T}{\alpha\nu}{\alpha}\right)\,.
\end{equation}
The superpotential plays an important role in representing TG as a gauge current for a gravitational energy-momentum tensor \cite{deAndrade:2000kr}. Contracting the torsion and superpotential tensors, the so-called torsion scalar is produced through
\begin{equation}
T := \dut{S}{a}{\mu\nu}\udt{T}{a}{\mu\nu}\,,
\end{equation}
which is determined solely by the Weitzenb\"{o}ck connection but can be used to compare with results in standard gravity. Coincidentally, it turns out that the torsion and Ricci scalars are equal up to a total divergence term \cite{Bahamonde:2017ifa,Bahamonde:2015zma}, namely
\begin{equation}\label{TEGR_L}
R=\mathring{R} +T-\frac{2}{e}\partial_{\mu}\left(e\udut{T}{\sigma}{\sigma}{\mu}\right)=0, \quad \Rightarrow \quad \mathring{R} = -T + \frac{2}{e}\partial_{\mu}\left(e\udut{T}{\sigma}{\sigma}{\mu}\right) := -T + B\,,
\end{equation}
where $\mathring{R}$ is the Ricci scalar as determined using the Levi-Civita connection, $R$ is the Ricci scalar as calculated with the Weitzeonb\"{o}ck connection which vanishes, and $e$ is the determinant of the tetrad field, $e = \det\left(\udt{e}{a}{\mu}\right) = \sqrt{-g}$. This relation alone guarantees that the torsion and Ricci scalars produce the same dynamical equations. Also, this means that the second- and fourth-order tetrad derivative contributions to the field equations can be somewhat decoupled in TG. In curvature-based theories, the Ricci scalar  couples these contributions together in a way prescribed by Eq.(\ref{TEGR_L}). This has important consequences for providing a more natural generalization  of $f(\mathring{R})$ gravity \cite{Capozziello:2018qcp}.

\noindent One straightforward result of this equivalency is that we can define TEGR as \cite{Bahamonde:2017wwk}
\begin{equation}
\mathcal{S}_{\rm TEGR} = -\frac{1}{2\kappa^2} \int d^4x \: e T + \int d^4x \: e \mathcal{L}_{\rm m}\,,
\end{equation}
where $\kappa^2 = 8\pi G$, and $\mathcal{L}_{\rm m}$ represents the Lagrangian for matter. Consequently, TEGR will produce identical Einstein field equations
\begin{equation}\label{TEGR_FEs}
    \mathring{G}_{\mu\nu} \equiv e^{-1}\udt{e}{a}{\mu}g_{\nu\rho}\partial_\sigma(e \dut{S}{a}{\rho\sigma})-\dudt{S}{b}{\sigma}{\nu}\udt{T}{b}{\sigma\mu}+\frac{1}{4}T g_{\mu\nu}-\udt{e}{a}{\mu} \udt{\omega}{b}{a\sigma}\dut{S}{b\nu}{\sigma} = \kappa^2 \Theta_{\mu\nu}\,,
\end{equation}
where $\Theta_{\mu\nu}$ is the energy-momentum tensor \cite{misner1973gravitation}, and $\mathring{G}_{\mu\nu}$ is the regular Einstein tensor calculated with the Levi-Civita connection.

Similar to the $f(\mathring{R})$ generalization of GR \cite{Faraoni:2008mf,Capozziello:2018qcp}, the TEGR Lagrangian density can be generalized to $f(T)$ gravity \cite{Ferraro:2006jd,Ferraro:2008ey,Bengochea:2008gz,Linder:2010py,Chen:2010va}. This produces second-order field equations \cite{Cai:2015emx}, as well as a number similarities to GR such as exhibiting the same gravitational wave polarizations \cite{Farrugia:2018gyz}. However, to incorporate both the second- and fourth-order components of $f(\mathring{R})$, it is advantageous to consider $\tilde{f}(T,B)$ gravity which forms a larger and richer class of theories than those expressed through $f(\mathring{R})$ gravity (at the level of field equations) \cite{Bahamonde:2015zma,Capozziello:2018qcp,Bahamonde:2016grb,Paliathanasis:2017flf,Farrugia:2018gyz,Bahamonde:2016cul,Bahamonde:2016cul,Wright:2016ayu,Farrugia:2020fcu}. By generalizing $f(\mathring{R})$ gravity in terms of its order contributions, $\tilde{f}(T,B)$ gravity may provide an interesting avenue to study fourth-order modified theories of gravity. As in theories based on the Levi-Civita connection, this will produce ten independent equations of motion that describe the dynamics of the system. However, a second set of six equations will also emerge through the anti-symmetric operator on the field equations. As investigated in Ref.\cite{Li:2010cg}, these equations must vanish due to the symmetry of the energy-momentum tensor on these indices. These extra equations represent the six LLTs. In general, it is these equations that are used to determine the spin connection components. It is for this reason that the tetrad and spin connection are so interrelated in that a choice in the spin connection components directly effects the equations that determine the tetrad components, and vice versa. The end result is that the teleparallel analog of the metric tensor is not only the tetrad but the tetrad together with its associated spin connection.

Given that $\tilde{f}(T,B)$ gravity produces fourth-order field equations, it is expected that if it produces Gauss-Ostrogradsky ghosts then they will disappear in the limit as $\tilde{f}(T,B)\Rightarrow \tilde{f}(-T+B)=f(\mathring{R})$ since $f(\mathring{R})$ gravity does not produce ghosts \cite{RevModPhys.82.451,DeFelice2010}. On the other hand, the models that we develop here exist already in the literature, and our aim is to probe their observational relevance in terms of whether they can confront current observation in the late-time Universe.


\section{$f(T,B)$ Cosmology}
\label{sec:f(T,B)_cosmo}

To explore the cosmology that emerges from $\tilde{f}(T,B)$ gravity, we consider a flat homogeneous and isotropic metric. We choose to take this FLRW metric in Cartesian coordinates so that it takes the form
\begin{equation}
ds^2=-dt^2+a(t)^2(dx^2+dy^2+dz^2)\,,
\end{equation}
where $a(t)$ is the scale factor, and the lapse function is already set to unity. This can be done since $\tilde{f}(T,B)$ gravity retains diffeomorphism invariance. By taking the choice of tetrad as \cite{Krssak:2018ywd}
\begin{equation}\label{FLRW_tetrad}
\udt{e}{a}{\mu}=\mbox{diag}(1,a(t),a(t),a(t))\,,
\end{equation}
the spin connection components are allowed to be zero, i.e. $\omega^{a}{}_{b\mu}=0$ \cite{Bahamonde:2016grb}. There exist an infinite number of possible choices for the tetrad which satisfy (\ref{Min_trans_Gen}) but only a small subset are good tetrads, i.e. have vanishing associated spin connection components. For this spacetime, the torsion scalar is explicitly given by
\begin{equation}\label{torsionscalar_frw}
T = 6H^2\,,
\end{equation}
while the boundary term is given by
\begin{equation}\label{boundaryscalar_frw}
B = 6\left(3H^2+\dot{H}\right)\,.
\end{equation}
Together, these form the Ricci scalar through the relation in Eq.(\ref{TEGR_L}), that is
\begin{equation}
\mathring{R} = -T+B = 6\left(\dot{H} + 2H^2\right)\,,
\end{equation}
is recovered. This shows how $f(\mathring{R})$ gravity results as a subset of $\tilde{f}(T,B)$ gravity where
\begin{equation}
\tilde{f}(T,B) := \tilde{f}(-T+B) = \tilde{f}(\mathring{R})\,,
\end{equation}
which only represents a small part of the space of models in $\tilde{f}(T,B)$ gravity. Also, the respectively second- and fourth-order contributions of the torsion scalar and boundary term can be seen directly through this choice of tetrad.

Evaluating the field equations for a Universe filled with a perfect fluid, the Friedmann equations turn out to be given by \cite{Bahamonde:2016cul,Bahamonde:2016grb}
\begin{eqnarray}
-3H^2\left(3f_B + 2f_T\right) + 3H\dot{f}_B - 3\dot{H} f_B + \frac{1}{2}f &=& \kappa^2\rho_m\label{Friedmann_1}\,,\\
-\left(3H^2+\dot{H}\right)\left(3f_B + 2f_T\right) - 2H\dot{f}_T + \ddot{f}_B + \frac{1}{2}f &=& -\kappa^2 p_m\label{Friedmann_2}\,,
\end{eqnarray}
where $\rho_m$ and $p_m$ represent the energy density and pressure of the matter content respectively. The Friedmann equations in Eqs.(\ref{Friedmann_1}-\ref{Friedmann_2}) show explicitly how a linear boundary contribution to the Lagrangian would act as a boundary term while other contributions of $B$ would contribute nontrivially to the dynamics of these equations. On taking the arbitrary Lagrangian mapping
\begin{equation}
\tilde{f}(T,B) \rightarrow -T + f(T,B)\,,
\end{equation}
the field equations can be re-expressed as
\begin{eqnarray}
3H^2 &=& \kappa^2 \left(\rho_m+\rho_{\mbox{eff}}\right)\,,\\
3H^2+\dot{H} &=& -\kappa^2\left(p_m+p_{\mbox{eff}}\right)\,,
\end{eqnarray}
where the modified TEGR components are contained in the effective fluid contributions given as
\begin{eqnarray}
\kappa^2 \rho_{\mbox{eff}} &=& 3H^2\left(3f_B + 2f_T\right) - 3H\dot{f}_B + 3\dot{H}f_B - \frac{1}{2}f\,, \label{eq:friedmann_mod}\\
\kappa^2 p_{\mbox{eff}} &=& \frac{1}{2}f-\left(3H^2+\dot{H}\right)\left(3f_B + 2f_T\right)-2H\dot{f}_T+\ddot{f}_B\,,
\end{eqnarray}
which can be combined to give
\begin{equation}
2\dot{H}=-\kappa^2\left(\rho_m + p_m + \rho_{\mbox{eff}} + p_{\mbox{eff}}\right)\,.
\end{equation}
The effective fluid that acts as the modified part of the $f(T,B)$ Lagrangian turns out to also satisfy the conservation equation
\begin{equation}
\dot{\rho}_{\mbox{eff}}+3H\left(\rho_{\mbox{eff}}+p_{\mbox{eff}}\right) = 0\,.
\end{equation}

\noindent Finally, an equation of state (EoS) can be written for this effective fluid as
\begin{eqnarray}
w_{\mbox{eff}} &=& \frac{p_{\mbox{eff}}}{\rho_{\mbox{eff}}}\\ 
&=& -1+\frac{\ddot{f}_B-3H\dot{f}_B-2\dot{H}f_T-2H\dot{f}_T}{3H^2\left(3f_B+2f_T\right)-3H\dot{f}_B+3\dot{H}f_B-\frac{1}{2}f}\,. \label{EoS_func}
\end{eqnarray}
Notice that we can recover the standard $\Lambda$CDM case ($w_{\mbox{eff}}=-1$) when we switch off the $T$ and $B$ terms. Since the latter equation is linked to a specific form
of $f(T,B)$, in this work we consider four cosmological models in order to investigate the possibility the effects of a late-time cosmic accelerated expansion without the influence of an
exotic dark energy or extra fields. It is important to remark that, in comparison to \cite{Cai:2015emx}, we solve the entire full system of equations (\ref{torsionscalar_frw})-(\ref{boundaryscalar_frw}) and the corresponding cosmological model.


\section{Current observational data}
\label{sec:data_analysis}

Given that we are interested in modelling the late-time evolution of the Universe, we use observational data from SNeIA luminosity distance from Pantheon compilation and BAO redshift surveys \cite{Scolnic:2017caz} and the high-z measurements of $H(z)$ from Galaxy ages (CC) \cite{Moresco:2016mzx}.

In order to carry out cosmological tests for the free parameters of each of the $f(T,B)$ models proposed below, we are going to consider the constraints solutions over $T$ and $B$ imposed by each case and determine the specific cosmological parameters for each model in addition to $\Omega_m$ and $H_0$ late Universe data. We use the publicly codes CLASS \footnote{\url{https://github.com/lesgourg/class_public}} and Monte Python \footnote{\url{https://github.com/baudren/montepython_public}} to constrain the models using a total sampler of CC+SNeIa+BAO.

\begin{figure}
\centering
\includegraphics[width=0.4\textwidth,origin=c,angle=0]{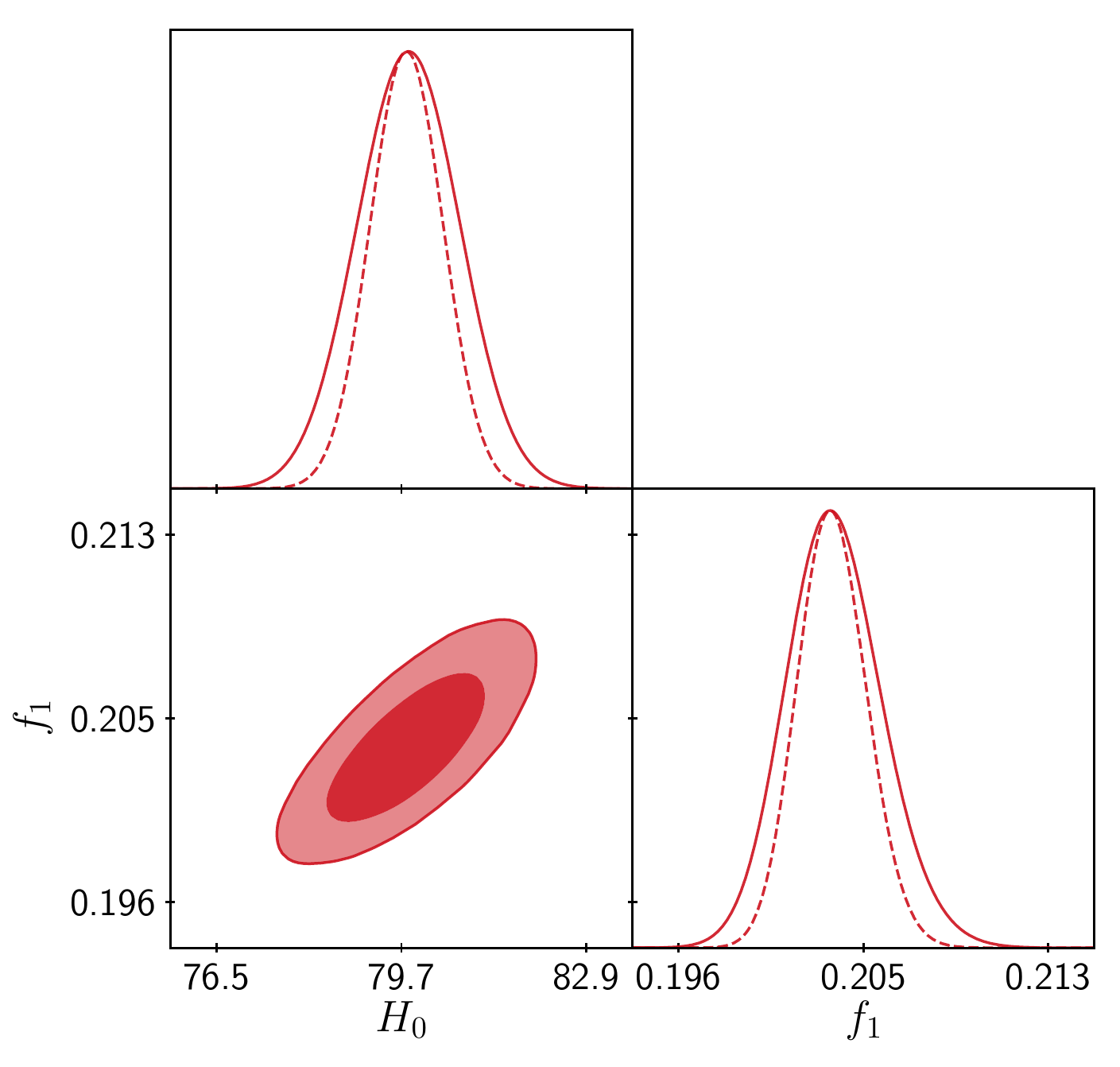}
\includegraphics[width=0.19\textwidth,origin=c,angle=0]{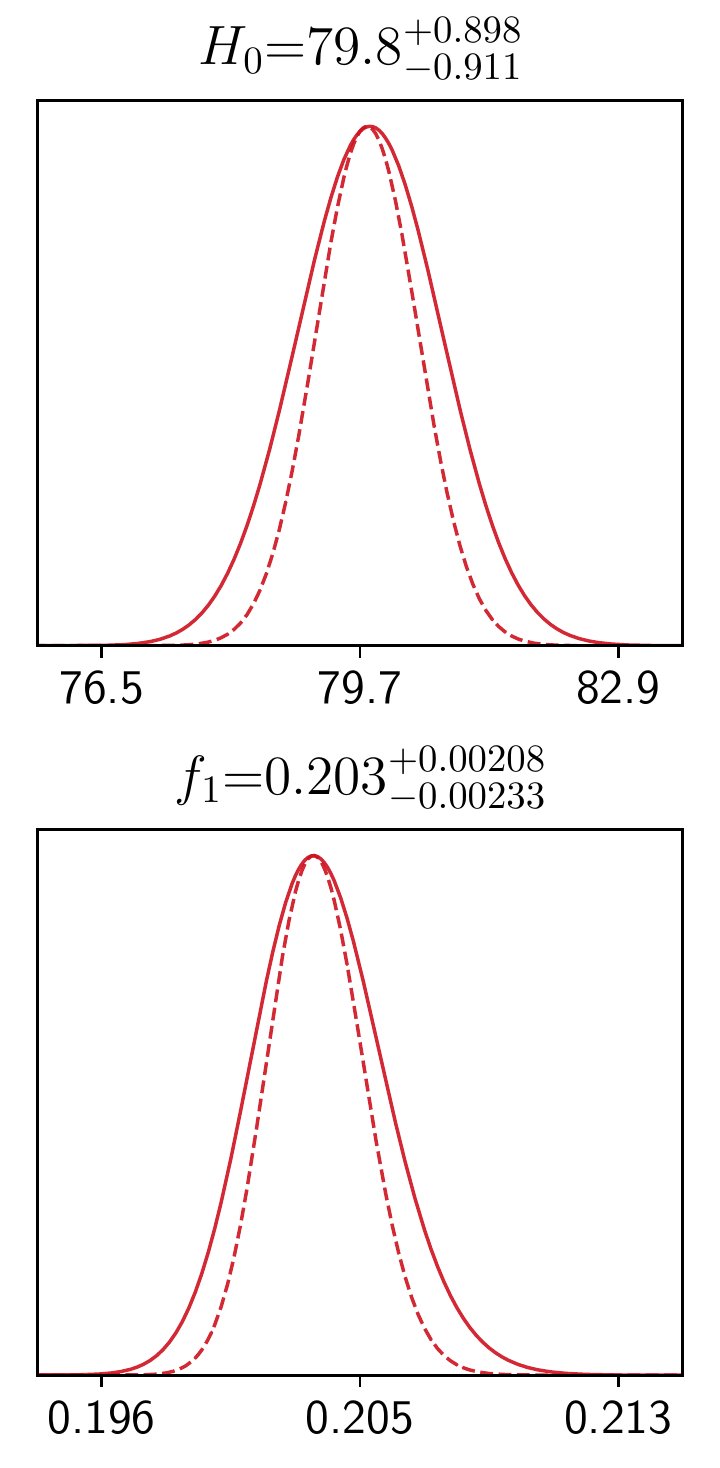}
\caption{One-dimensional marginalised distribution, and two-dimensional contours with $68\%$ and $95\%$ confidence level for the free parameters for TEGR Deviations model contours using the constrained solutions for $T$ and $B$ scalars and CC+Pantheon+BAO total sampler.} 
\label{contour_TEGRcontours}
\end{figure}

Figures \ref{contour_taylor_case1}-\ref{contour_powerlaw}-\ref{contour_mixedlaw}--\ref{contour_TEGRcontours} show the parameter space for $w_{\mbox{model}}$ and $\Omega_m$ with their probability density function (PDF) versus $\Omega_m$ up to $3$-$\sigma$ confidences levels (CL) using the joint sampler CC+Pantheon+BAO.


\subsection{Galaxy ages}

Given that distance scale measurements require integrals of $H(z)$, it is a standard point of view that it is more precise to study the observational $H(z)$ data directly rather than these means since information loses are a natural consequence of these integrals, and of course, the errors that these can carry out. As an independent approach of this measure we provide the Cosmic Chronometers (CC) sample. This kind of sample gives a measurement of the expansion
rate without relying on the nature of the metric between the chronometer and us. A full compilation of the latter, which includes 38 measurements
of $H(z)$ in the range $ 0.07 < z < 2.36$ are reported in \cite{Moresco:2016mzx}. 

\subsection{Pantheon Type Ia supernovae compilation}
In \cite{Scolnic:2017caz}  was presented a SNeIa sampler with 1080 data points compressed in 40 bins. This binned catalog is in agreement with the standard 
$\Lambda$CDM EoS. We start by defining the distance modulus $\mu$ in relation to the luminosity distance $d_L$ in Mpc as:
\begin{equation}\label{eq:lum}
\mu(z)= 5\ln{\left[\frac{d_L (z)}{1 \mbox{Mpc}}\right]} +25\,,
\end{equation}
Since we are considering the hypothesis of spatial flatness, the luminosity distance can be described using the comoving distance $D$ as
\begin{equation}
d_{L} (z) =\frac{c}{H_0} (1+z)D(z)\,,
\end{equation}
where $c$ is the speed of light. Using the latter we can write  
\begin{equation}
D(z) =\frac{H_0}{c}(1+z)^{-1}10^{\frac{\mu(z)}{5}-5}\,. \label{eq:D}
\end{equation}
The normalised Hubble function $E=H(z)/H_0$ can be obtained by taking the inverse of the derivative of (\ref{eq:D}) with respect to the redshift  as: $D(z)=\int^{z}_{0} H_0 d\tilde{z}/H(\tilde{z})$.
Also, we are going to consider uniform priors based to Planck value in Eq.(\ref{eq:planck2018}) with TT, TE, EE + lowE + lensing +BAO and to the Late Universe measurements as
\begin{eqnarray}
H_0 &=&67.4 \pm 0.5 \mbox{km/s/Mpc}, \quad \mbox{from Planck 2018}, \label{eq:planck2018}\\
H_0 &=&73.8 \pm 1.1 \mbox{km/s/Mpc},  \quad \mbox{from Late Universe (SH0ES + H0LiCOW)} \label{eq:SH0ES+H0LiCOW},
\end{eqnarray}
We used a Monte Python code and to obtain the following values for the nuisance parameter: $M=-19.63$ for Planck 2018 and $M=-32.79$ for SH0ES+H0LiCOW.

\subsection{BAO samples}
We also consider in our analysis the measurements of BAO observations
in the galaxy distribution. These observations can contribute important features by comparing the data of the
sound horizon today to the sound horizon at the time of recombination (extracted from the CMB anisotropy~data).
Commonly, the BAO distances are given as a combination of the angular scale and the \mbox{redshift separation}:
$d_{z} \equiv \frac{r_{s}(z_{d})}{D_{V}(z)}$, with  $r_{s}(z_{d}) = \frac{c}{H_{0}} \int_{z_{d}}^{\infty}
\frac{c_{s}(z)}{E(z)} \mathrm{d}z$ and
$r_{s}(z_{d})$ being the comoving sound horizon at the baryon dragging epoch,
$c$ the light velocity, $z_{d}$ is the
drag epoch redshift and $c^{2}_{s}= c^2/3[1+(3\Omega_{b0}/4\Omega_{\gamma 0})(1+z)^{-1}]$ the sound speed
with $\Omega_{b0}$ and $\Omega_{\gamma 0}$ the present values of baryon and photon density parameters, respectively. By definition, the dilation scale is
\begin{equation}
D_{V}(z,\Omega_m; w_0,w_1) = \left[ (1+z)^2 D_{A}^2 \frac{c
\, z}{H(z, \Omega_m; w_0,w_1)} \right]^{1/3}\,,
\end{equation}
where $D_{A}$ is the angular diameter distance
\begin{equation}
D_{A}(z,\Omega_m; w_0,w_1) = \frac{1}{1+z} \int_{0}^{z}
\frac{c \, \mathrm{d}\tilde{z}}{H(\tilde{z}, \Omega_m;w_0,w_1)} \,.
\end{equation}

Through the comoving sound horizon, the distance ratio $d_{z}$ is
related to the expansion parameter $h$ (defined such that {$H \doteq 100 h$)} and the physical densities $\Omega_{m}$ and $\Omega_{b}$.
We use  measurements of the BAO peak from the galaxy redshift surveys six-degree-field galaxy survey (6dFGS, 
Sloan Digital Sky Survey Data Release 7 (SDSS DR7) and the reconstructed value (SDSS(R)), as well as the latest result from the complete BOSS sample SDSS DR12, and also from the Lyman-$\alpha$ Forest measurements from the Baryon Oscillation Spectroscopic Data Release 11 (BOSS DR11).  
Since the volume surveyed by BOSS and WiggleZ \cite{Kazin:2014qga} partially overlap we do not use data from the latter in this work (see details in Ref.\cite{Beutler:2015tla}).
The total $\chi^2_{\mathrm{BAO}}$ is directly obtained by the sum of the individual quantity:
$\chi^2_{\mathrm{BAO-total}}=\chi^2_{\mathrm{6dFGS}} +\chi^2_{\mathrm{SDSS}} +\chi^2_{\mathrm{BOSS}} +\chi^2_{\mathrm{Ly}\alpha\mathrm{-F}}$. The full sampler of these data is shown in Table \ref{Table:bao}.

\begin{table}
\begin{center}
\begin{tabular} { llcclll }
\hline
Data set      & \,\,\,\, &$z$ \,\,\, & $r_{\mbox{BAO}}(z)$ &  \\
\hline
 6dF  			       & \,\,\,\, & $0.106$ \cite{Beutler:2011hx}           &$0.336\pm 0.015$ \\    		          
 SDSS DR7			       & \,\,\,\, & $0.15$ \cite{Howlett:2014opa}            &$0.2239\pm 0.0084$ \\      
  SDSS(R) DR7 			       & \,\,\,\, & $0.35$ \cite{Mehta:2012hh}          &$0.1137\pm 0.0021$ \\     
   SDSS(R)-III DR12 			       & \,\,\,\, & $0.38$  \cite{Stoppacher:2019ssr}          &$0.100\pm 0.0011$ \\   
   			       & \,\,\,\, & $0.61$     \cite{Alam:2016hwk}       &$0.0691\pm 0.0007$ \\    
      SDSS(R)-III DR12 			       & \,\,\,\, & $2.34$    \cite{Delubac:2014aqe}       &$0.0320\pm 0.0013$ \\
    			       & \,\,\,\, & $2.36$  \cite{Font-Ribera:2013wce}          &$0.0329\pm 0.0009$ \\
\hline
\end{tabular}
\caption{$r_{\mbox{BAO}}(z)$ measurements used in this work. The selected ones corresponding to SDSS data were inverted from the published values
of $D_V(z)/s_d$ and those corresponding to Ly$\alpha$-F data were obtained from the reported quantities $D_A(z)/s_d$ and $D_H(z)/s_d$.}
\label{Table:bao}
\end{center}
\end{table}


\section{Cosmologically inspired $f(T,B)$ models}
\label{sec:models}
In this section we are going to consider four possible cosmological models in order to study late time cosmic accelerations
and compliments this with a study of the effects of dynamical dark energy-like equation of state that each model produces.

\subsection{General Taylor Expansion Model}
As in \cite{Farrugia:2018gyz}, first consider a general Taylor expansion of the $f(T,B)$ Lagrangian, given as
\begin{eqnarray}
f(T,B) &=& f(T_0, B_0) + f_T(T_0,B_0) (T-T_0) + f_B(T_0,B_0) (B-B_0) \nonumber\\
 && + \frac{1}{2!}f_{TT}(T_0,B_0) (T-T_0)^2 + \frac{1}{2!}f_{BB}(T_0,B_0) (B-B_0)^2  \nonumber\\
 &&  + f_{TB}(T_0,B_0) (T-T_0)(B-B_0) + \mathcal{O}(T^3,B^3)\,,
\end{eqnarray}
where we need to go beyond linear approximations since $B$ is a boundary term at linear order. The FLRW tetrad in Eq.(\ref{FLRW_tetrad}) describes spacetime on cosmological scales, while locally spacetime appears to be Minkowski with torsion scalar and boundary term values
\begin{equation}
  T_0 = 0\,, \quad B_0 = 0\,.
\end{equation}
We expand about these local values to produce a general Lagrangian in which to study cosmology using the FLRW tetrad. It is not clear what the values of the arbitrary Lagrangian $f$ will be, and so taking constants $A_i$, the Lagrangian can be written as
\begin{equation}
f(T,B)\simeq A_0+A_1 T + A_2 T^2 + A_3 B^2 + A_4 TB\,,
\end{equation}
where the linear boundary term has been removed. All the derivatives corresponding to $f_T$ and $f_B$ functions for this model can be found in Appendix \ref{app:GTE}.

From Eq.(\ref{eq:friedmann_mod}) adding $3H^2$ to both sides we then write
\begin{equation}
H^2= \frac{1}{2}\left[3H^2 (1-3f_B -2f_T) +\frac{1}{2}f +3H \dot{f_B} -3\dot{H}f_B\right] +\frac{\kappa^2}{3}\rho\,,
\end{equation}
which can be comparing to $H^2=\kappa^2/3 (\rho+\rho_X)$, where $\rho_X$ is denominates the \textit{X-fluid}. We can define the terms in brackets as
\begin{equation}
\kappa^2 \rho_X := 3H^2 (1-3f_B -2f_T)+\frac{1}{2}f +3H \dot{f_B} -3\dot{H}f_B\,,
\end{equation}
in this way, the matter term does not appear explicitly in $\rho_X$. For this case, the corresponding EoS is given by \footnote{We represented here $a^{(i)}$, with $i>2$ as high derivatives with respect to $t$.}
\begin{eqnarray}
 &&w_{x}(a)=\nonumber\\
 &&\frac{-24w_{x_7} \dot{a}(t)^5+6 a(t)^4 \left[\ddot{a}(t)-4 A_3 a^{(4)}(t)\right]+24 a(t) \dot{a}(t)^3 w_{x_1} 
 -6 a(t)^3 w_{x_2}
 +24 a(t)^2 \dot{a}(t) w_{x_3}
   +w_{x_8}}
   {-6 a(t)^4 \ddot{a}(t)
   +w_{x_5}
   +72 a(t) \dot{a}(t)^3  w_{x_4}
   +w_{x_6}}\,,\nonumber\\
   &&
\end{eqnarray}
where
\begin{eqnarray}
w_{x_1}&=& \left[\left(4 A_2+27 A_3+11 A_4\right) \ddot{a}(t)-\left(4 A_2+12 A_3+7 A_4\right) \dot{a}(t)\right]\,, \\
w_{x_2}&=& \left[4\left(3 A_3+A_4\right) \ddot{a}(t)^2+\dot{a}(t)^2+4 a^{(3)}(t) \left(2 A_3 \dot{a}(t)-A_4 \ddot{a}(t)\right)\right]\,, \\
w_{x_3}&=& \left[\left(9 A_3+2 A_4\right) a^{(3)}(t) \dot{a}(t)+\ddot{a}(t) \left(\left(4 A_2+3 A_4\right) \ddot{a}(t)+2 \left(2 A_2+9
   A_3+4 A_4\right) \dot{a}(t)\right)\right]\,, \\
   w_{x_4}&=& -\left[\left(4 A_2+9 A_3+6 A_4\right) \ddot{a}(t)+\left(4   A_3+A_4\right) \dot{a}(t)\right]\,, \\
   w_{x_5}&=& 72 \left(4 A_2+12 A_3+7 A_4\right) \dot{a}(t)^5+6 a(t)^3 \dot{a}(t)\left(12 A_3 a^{(3)}(t)+\dot{a}(t)\right)\,, \\
   w_{x_6}&=& 72 \left(3 A_3+A_4\right) a(t)^2 \dot{a}(t)^2 \left[\ddot{a}(t)-a^{(3)}(t)\right]+a(t)^5\left[A_3 B^2+T \left(A_4 B+A_2 T+A_1\right)+A_0\right]\,, \nonumber\\
   &&\\
   w_{x_7}&=&  8 A_2+36 A_3+17 A_4\,, \\
   w_{x_8}&=& a(t)^5 \left[-\left(A_3 B^2+T \left(A_4 B+A_2T+A_1\right)+A_0\right)\right]\,,
\end{eqnarray}

\begin{figure}
\centering
\includegraphics[width=0.32\textwidth,origin=c,angle=0]{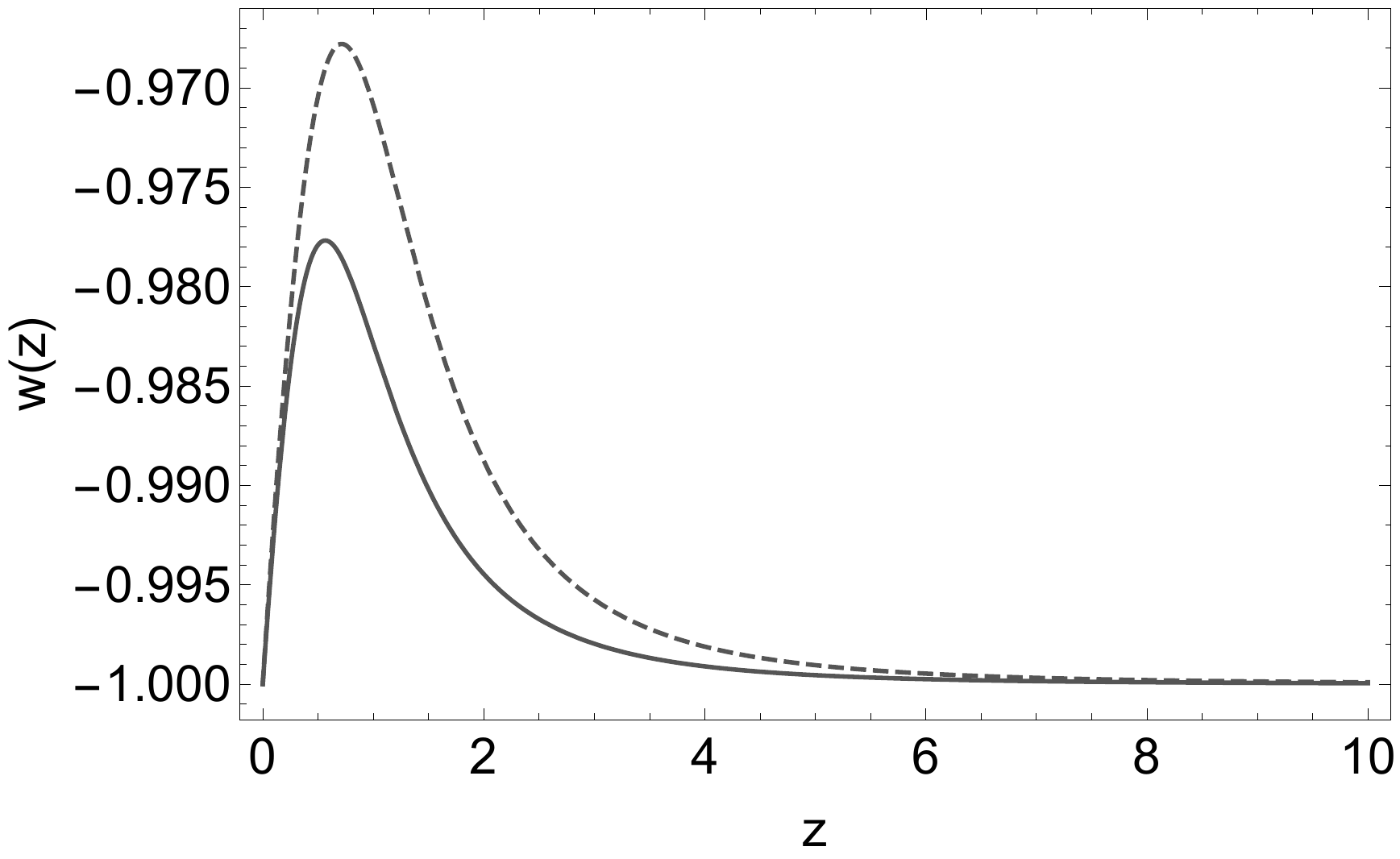}
\includegraphics[width=0.31\textwidth,origin=c,angle=0]{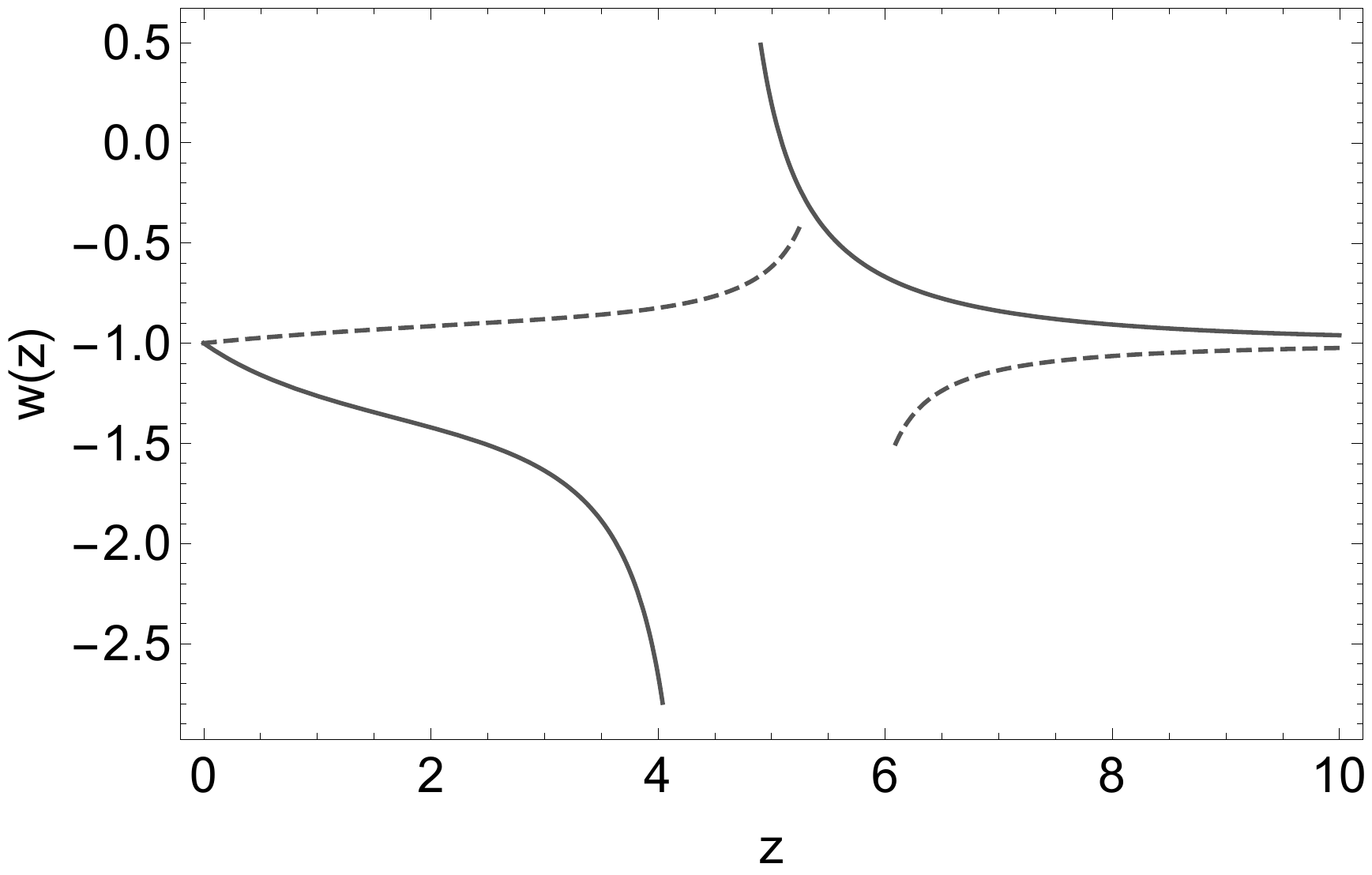}
\includegraphics[width=0.31\textwidth,origin=c,angle=0]{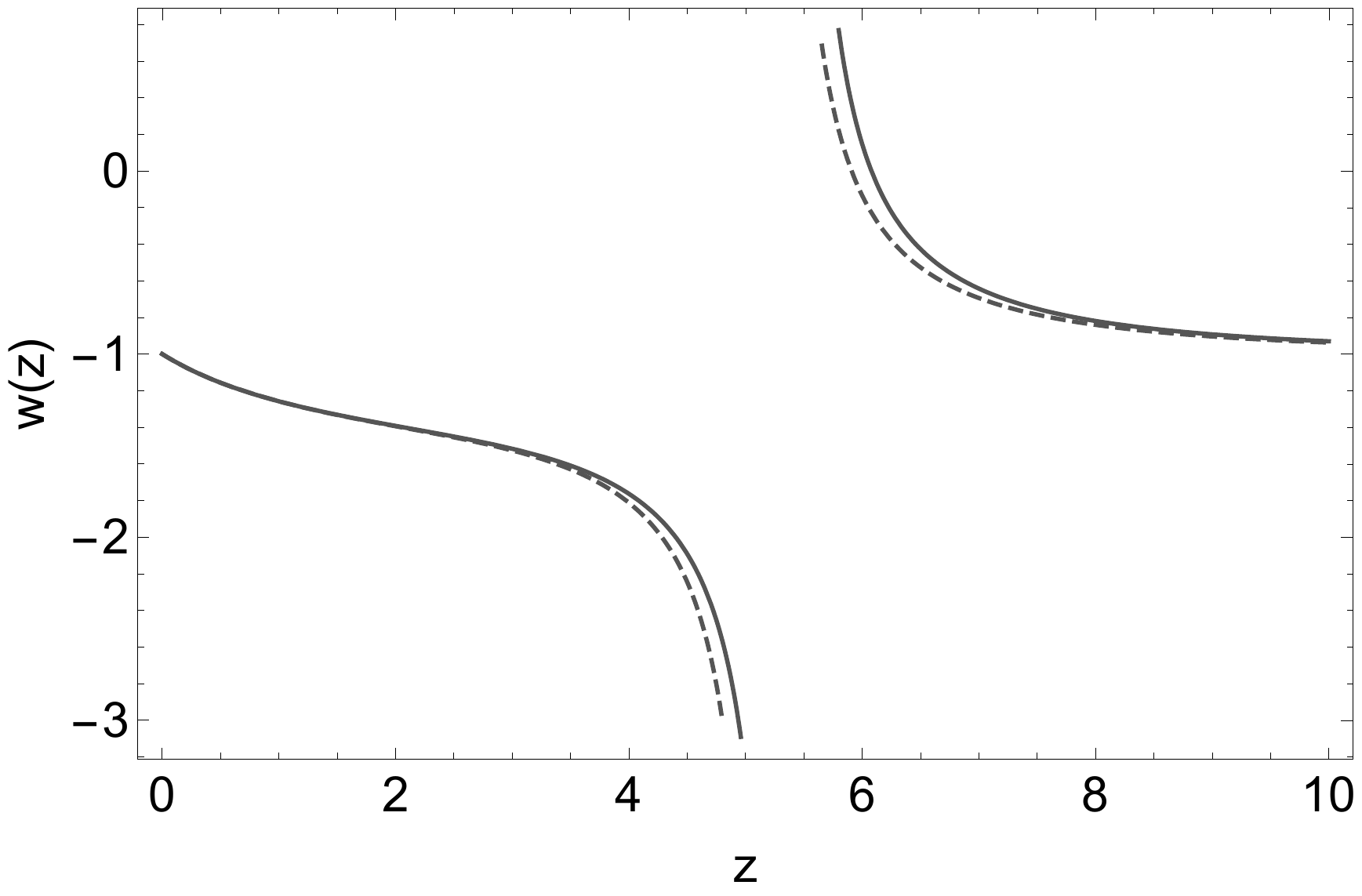}
\caption{Evolution of General Taylor Expansion EoS (\ref{eq:eosz_taylor}). \textit{Left:} Case 1: with the condition that $T<B$ (solid line) and $T>B$ (dashed line). \textit{Middle:} Case 2.1: $A_{i+1} > A_i$ with $T<B$ (solid line) and $T>B$ (dashed line) . \textit{Right:} Case 2.2: $A_{i+1} > A_i$ with $T<B$ (solid line) and $T>B$ (dashed line).} 
\label{evolution_taylor}
\end{figure}

In order to perform the numerical analysis, we rewrite the above expression in terms of redshift $z=a_0 /a -1$, where $z=0$ corresponds
to the present time. The EoS of this model can be expressed as
\begin{eqnarray} \label{eq:eosz_taylor}
w(z)&=& \frac{w(z)_1
+6\left[w(z)_2 -w(z)_6\right]
+24\left[w(z)_3-w(z)_4 -w(z)_5\right]}
   {-w(z)_1
   -6 w(z)_7
   +72 \left[w(z)_8- w(z)_9 -w(z)_{10}\right]
   -\frac{12}{(z+1)^7}}\,,
\end{eqnarray}
where each  $w(z)_{i}$ function is given by
\begin{eqnarray}
w(z)_1&=& -\frac{A_3 B^2+T \left(A_4 B+A_2 T+A_1\right)+A_0}{(z+1)^5}\,, \\
w(z)_2&=& \frac{\frac{2}{(z+1)^3}-\frac{96A_3}{(z+1)^5}}{(z+1)^4}\,,\\
w(z)_3&=& \frac{\left(8 A_2+36 A_3+17 A_4\right)}{(z+1)^{10}}\,, \\
w(z)_4&=& \frac{\frac{2 \left(4A_2+27 A_3+11 A_4\right)}{(z+1)^3}+\frac{4 A_2+12 A_3+7 A_4}{(z+1)^2}}{(z+1)^7}\,, \\
w(z)_5&=&  \frac{\frac{6 \left(9 A_3+2A_4\right)}{(z+1)^6}+\frac{2 \left(\frac{2 \left(4 A_2+3 A_4\right)}{(z+1)^3}-\frac{2 \left(2 A_2+9 A_3+4
   A_4\right)}{(z+1)^2}\right)}{(z+1)^3}}{(z+1)^4}\,, \\
   w(z)_6&=& \frac{\frac{16 \left(3A_3+A_4\right)}{(z+1)^6}-\frac{24 \left(-\frac{2 A_3}{(z+1)^2}-\frac{2
   A_4}{(z+1)^3}\right)}{(z+1)^4}+\frac{1}{(z+1)^4}}{(z+1)^3}\,, \\
   w(z)_7&=&- \frac{\frac{72 A_3}{(z+1)^4}+\frac{1}{(z+1)^2}}{(z+1)^5}\,, \\
   w(z)_8&=& \frac{\left(3A_3+A_4\right) \left[\frac{2}{(z+1)^3}+\frac{6}{(z+1)^4}\right]}{(z+1)^6}\,, \\
   w(z)_9&=& \frac{4 A_2+12 A_3+7A_4}{(z+1)^{10}}\,,\\
   w(z)_{10}&=& \frac{\frac{4 A_3+A_4}{(z+1)^2}-\frac{2 \left(4 A_2+9 A_3+6A_4\right)}{(z+1)^3}}{(z+1)^7}\,.
\end{eqnarray}
To solve the system of equations in Eqs.(\ref{torsionscalar_frw})-(\ref{boundaryscalar_frw}) we are going to analyse the behaviour of the (\ref{eq:eosz_taylor}) EoS associated with the $X-$fluid considering four kind of cases (c.f. Figure \ref{evolution_taylor})
\begin{itemize}
\item Case 1.1: Solving the system Eqs.(\ref{torsionscalar_frw})-(\ref{boundaryscalar_frw}) with the condition that $T<B$, i.e  we have domination of the boundary term over the torsion scalar. Also we consider $A_i$ as positive values.
\item Case 1.2: Solving the system Eqs.(\ref{torsionscalar_frw})-(\ref{boundaryscalar_frw}) with the condition that $T>B$, i.e we have domination of the torsion scalar over the boundary term). Also we consider $A_i$ as positive values.
\item Case 2.1: Solving the system Eqs.(\ref{torsionscalar_frw})-(\ref{boundaryscalar_frw}) with the condition that $T<B$, i.e  we have domination of the boundary term over the torsion scalar. Also we consider $A_{i+1} > A_i$.
\item Case 2.2: Solving the system Eqs.(\ref{torsionscalar_frw})-(\ref{boundaryscalar_frw}) with the condition that $T>B$, i.e we have domination of the torsion scalar over the boundary term). Also we consider $A_{i+1} < A_i$.
\end{itemize}

Following the above scenarios (cf. with Figure \ref{evolution_taylor}), we notice that in both Cases 1.1 and 1.2 (cf. with Figure \ref{evolution_taylor} - \textit{left})  the General Taylor Expansion EoS mimic the same evolution at high redshifts, which indicates that at early times the boundary term and the torsion scalar are indistinguishable if we consider the influence of an \textit{X-fluid}. On the other hand, at $z\approx 1$ \footnote{Assuming a flat universe with $H_0 = 67.3 \mbox{km}\mbox{s}^{-1}\mbox{Mpc}^{-1}$, this redshift correspond to a look-back time of 7.95 Gyr.} a scenario where the boundary scalar dominate we have a quintessence-like behaviour that approaches more to a $\Lambda$CDM Eos than the scenario where the torsion scalar dominates. Cases 2.1 and 2.2  (cf. with Figure \ref{evolution_taylor} - \textit{middle} and \textit{right}) show a divergence due the degeneracies of the constants $A_{i}$.

\begin{figure}
\centering
\includegraphics[width=0.3\textwidth,origin=c,angle=0]{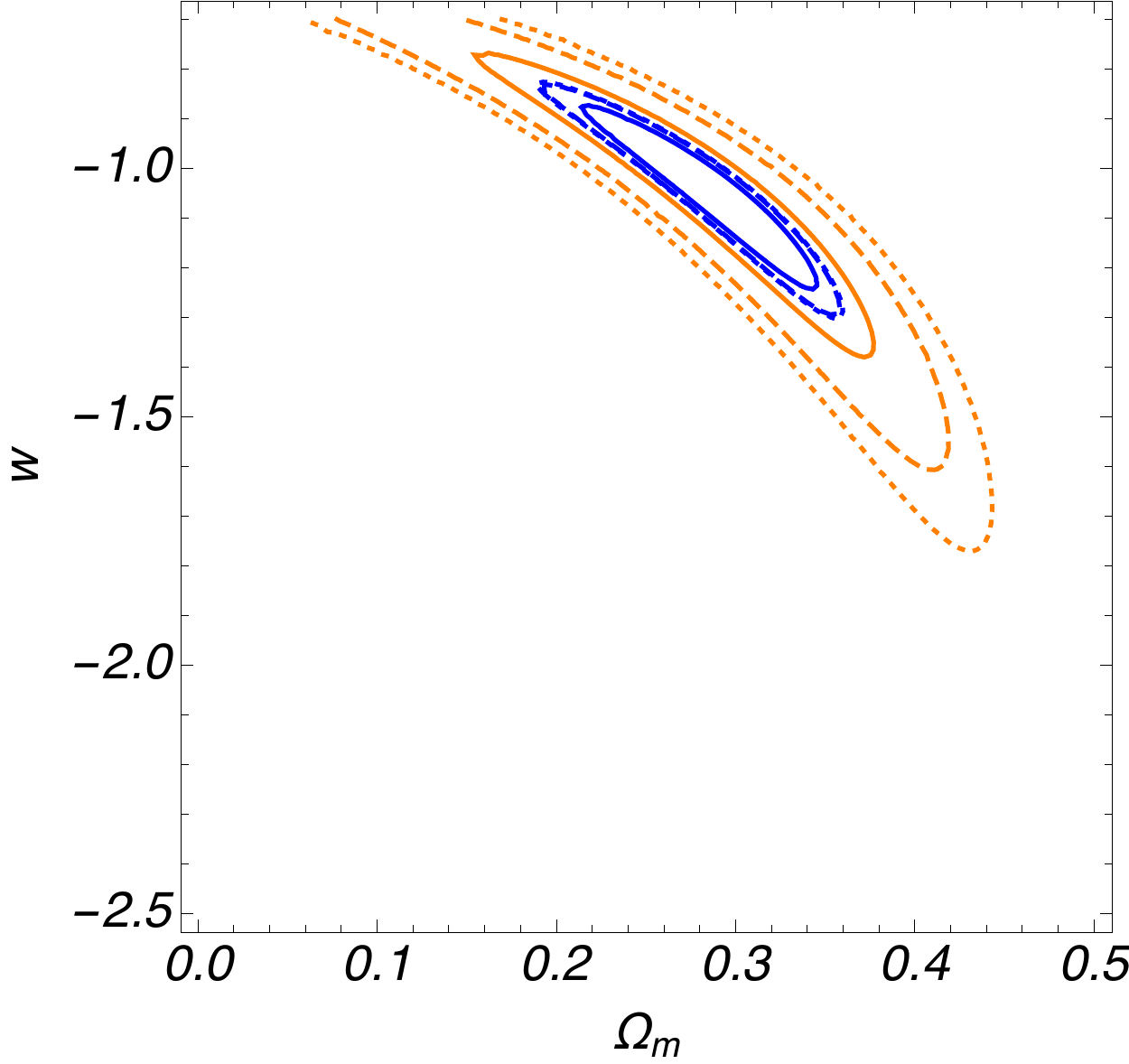}
\includegraphics[width=0.288\textwidth,origin=c,angle=0]{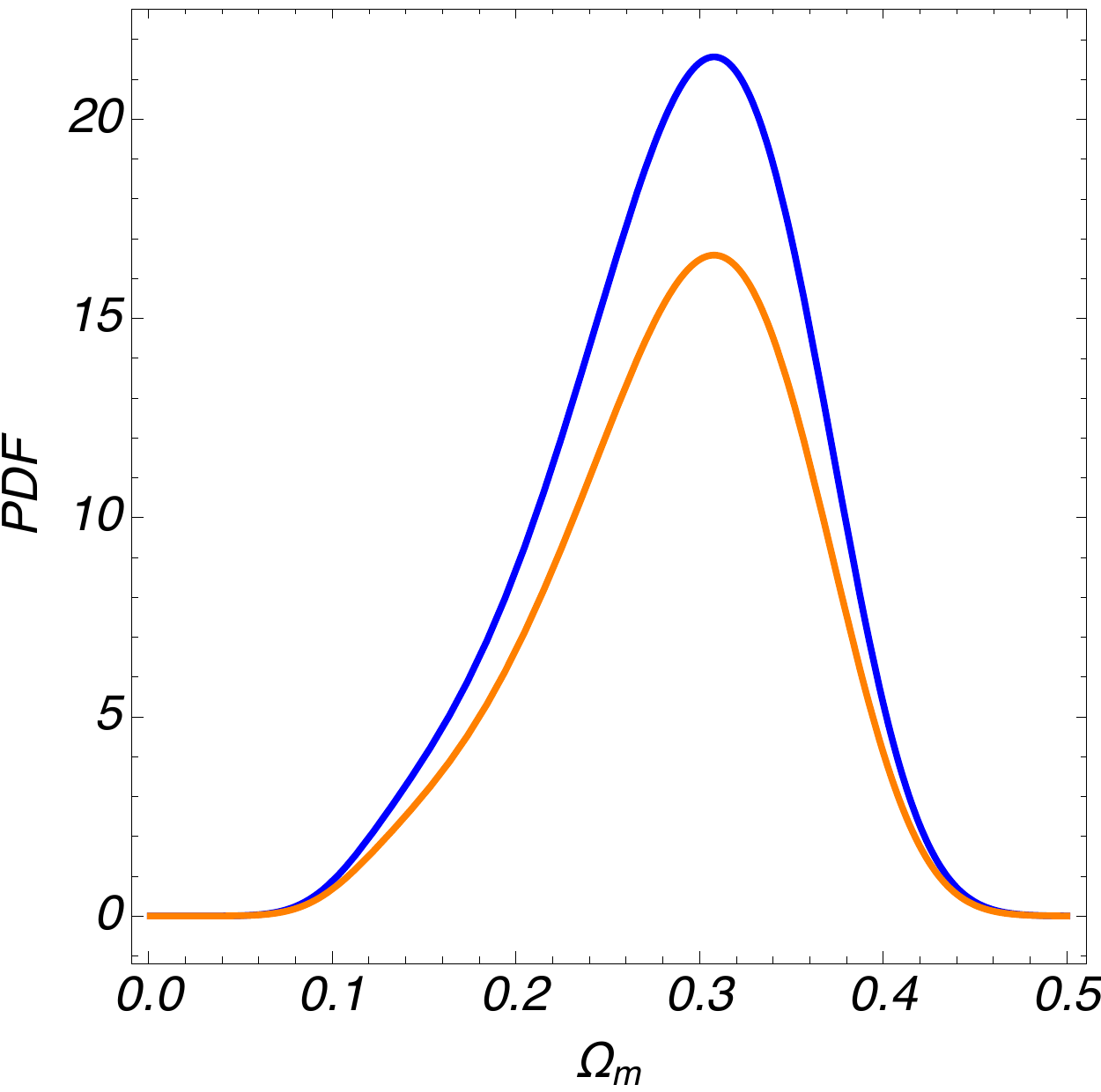} 
\caption{General Taylor Expansion model contours plots for Cases 1.1 (orange color) and 1.2 (blue color) using CC+Pantheon+BAO samplers.}
\label{contour_taylor_case1}
\end{figure}

For the model given in Eq.(\ref{eq:eosz_taylor}), we perform the fitting using the completed compilation of data samplers described in \S.~\ref{sec:data_analysis}. This analysis is showed in Figure \ref{contour_taylor_case1}, where we notice that the model has a preference for a phantom-like behaviour in agreement with Planck data value for the density of matter.

\subsection{Power Law Model}
Following Ref.\cite{Bahamonde:2016grb}, consider a Lagrangian of separated power law style models for the torsion and boundary scalars such that
\begin{equation}
f(T,B) = b_0 B^k + t_0 T^m\,. \label{eq:powerlaw}
\end{equation}

All the derivatives corresponding to $f_T$ and $f_B$ functions for this model can be found in the Appendix \ref{app:PL}.
Since we are interested in understanding whether this power law model can reproduce a dark energy-like behaviour, we compute the EoS in Eq.(\ref{EoS_func}) for the model in Eq.(\ref{eq:powerlaw}) and obtain
\begin{equation}
 w_{x}(a)=-1+\frac{\frac{b_0 6^k (k-1) k \left[\frac{a(t) \ddot{a}(t)+2 \dot{a}(t)^2}{a(t)^2}\right]^k w_{x_1}}
   {\left[a(t) \ddot{a}(t)+2 \dot{a}(t)^2\right]^3}-\frac{t_0 2^{m+2} 3^m (m-1) m a(t)
   \left[\frac{\dot{a}(t)^2}{a(t)^2}\right]^{m+1} w_{x_2}}{\dot{a}(t)^5}}
   {3 \left\{\frac{6 \left[a(t) \ddot{a}(t)-\dot{a}(t)^2\right]}{a(t)^2}+ w_{x_4}
   +w_{x_5}-b_0 B^k-t_0 T^m\right\}}\,,
\end{equation}
where the functions $w_{x_i}$ are given by
\begin{eqnarray}
w_{x_1}&=&\left\{(k-1) a(t)^5 \dddot{a}(t)+48 \dot{a}(t)^6-30 a(t)
  \dot{a}(t)^4 \ddot{a}(t)+a(t)^4 \ddot{a}(t) \left[\multidots{4}{a}(t)+3 (k-1) \dot{a}(t)\right]
  \right. \nonumber \\ && \left. -a(t)^2 \dot{a}(t)^2 \left[21 \ddot{a}(t)^2+2 \dddot{a}(t)
   \dot{a}(t)\right]+a(t)^3 \left[3 \ddot{a}(t)^3+\dot{a}(t) \left(2 \dot{a}(t) \left(\multidots{4}{a}(t)-2 (k-1) \dot{a}(t)\right)
   \right.\right.\right. \nonumber\\&& \left.\left.\left.
   -\dddot{a}(t) \ddot{a}(t)\right)\right]\right\}\,, \\
   w_{x_2}&=& \left[\dot{a}(t)^2-a(t) \ddot{a}(t)\right] \left\{\dot{a}(t)^2-a(t)\left[\ddot{a}(t)+\dot{a}(t)\right]\right\}\,, \\
    w_{x_3} &=& b_0 3^{k+1} (k-1) k \dot{a}(t)^3 \left[a(t)^2
   \dddot{a}(t)-4 \dot{a}(t)^3+3 a(t) \dot{a}(t) \ddot{a}(t)\right] \left[\frac{2 a(t) \ddot{a}(t)+4 \dot{a}(t)^2}{a(t)^2}\right]^k\,, \\
   w_{x_4}&=& \frac{t_0 2^{m+2} 3^m
   (m-1) m \left[\frac{\dot{a}(t)^2}{a(t)^2}\right]^m \left[a(t) \ddot{a}(t)-\dot{a}(t)^2\right]
   +\frac{w_{x_3}}
   {\left[a(t) \ddot{a}(t)+2
   \dot{a}(t)^2\right]^2}}{a(t) \dot{a}(t)}\,, \\
   w_{x_5}&=& \frac{b_0 6^k (k-1) k \dot{a}(t) \left[-a(t)^2 \dddot{a}(t)+4 \dot{a}(t)^3-3 a(t) \dot{a}(t) \ddot{a}(t)\right]
   \left[\frac{a(t) \ddot{a}(t)+2 \dot{a}(t)^2}{a(t)^2}\right]^k}{\left[a(t) \ddot{a}(t)+2 \dot{a}(t)^2\right]^2}\,.
\end{eqnarray}
To perform the numerical analysis, we rewrite the above expression in terms of the redshift $z=a_0 /a -1$, where $z=0$ corresponds
to the present time. The power law model EoS can be expressed as
\begin{eqnarray}\label{eq:eosz_powerlaw}
&&w(z)=-1\nonumber\\
&&+\frac{b_0 3^k 8^{k-2} (k-1) k (z+1)^{12} \left[\frac{1}{(z+1)^2}\right]^k 
w(z)_1
   -t_0 2^{m+2} 3^m
   (m-1) m (z+1)^5 
   w(z)_2
   \left[\frac{1}{(z+1)^2}\right]^{m+1}}{3 \left\{-b_0 B^k-(z+1)^3 
   w(z)_3
   -b_0 2^{3
   k-1} 3^k (k-1) k \left[\frac{1}{(z+1)^2}\right]^k-t_0 T^m+\frac{6}{(z+1)^2}\right\}}\,,\nonumber\\
   &&
\end{eqnarray}
where
\begin{eqnarray}
w(z)_1&=&\left\{-\frac{6 (k-1)}{(z+1)^9}+\frac{2
   \left[\frac{24}{(z+1)^5}-\frac{3 (k-1)}{(z+1)^2}\right]}{(z+1)^7}+\frac{\frac{24}{(z+1)^9}-\frac{\frac{12}{(z+1)^7}-\frac{2
   \left[\frac{2 (k-1)}{(z+1)^2}+\frac{24}{(z+1)^5}\right]}{(z+1)^2}}{(z+1)^2}}{(z+1)^3}-\frac{108}{(z+1)^{12}}\right\}\,, \nonumber\\
   &&\\
w(z)_2&=& \left[\frac{1}{(z+1)^4}-\frac{\frac{2}{(z+1)^3}-\frac{1}{(z+1)^2}}{z+1}\right]\,, \\
w(z)_3&=& \left\{\frac{b_0 2^{3 k-1} 3^{k+1} (k-1) k \left[\frac{1}{(z+1)^2}\right]^k}{(z+1)^4}+\frac{t_0 2^{m+2} 3^m (m-1) m \left[\frac{1}{(z+1)^2}\right]^m}{(z+1)^4}\right\}\,.
\end{eqnarray}

\begin{figure}
\centering
\includegraphics[width=0.32\textwidth,origin=c,angle=0]{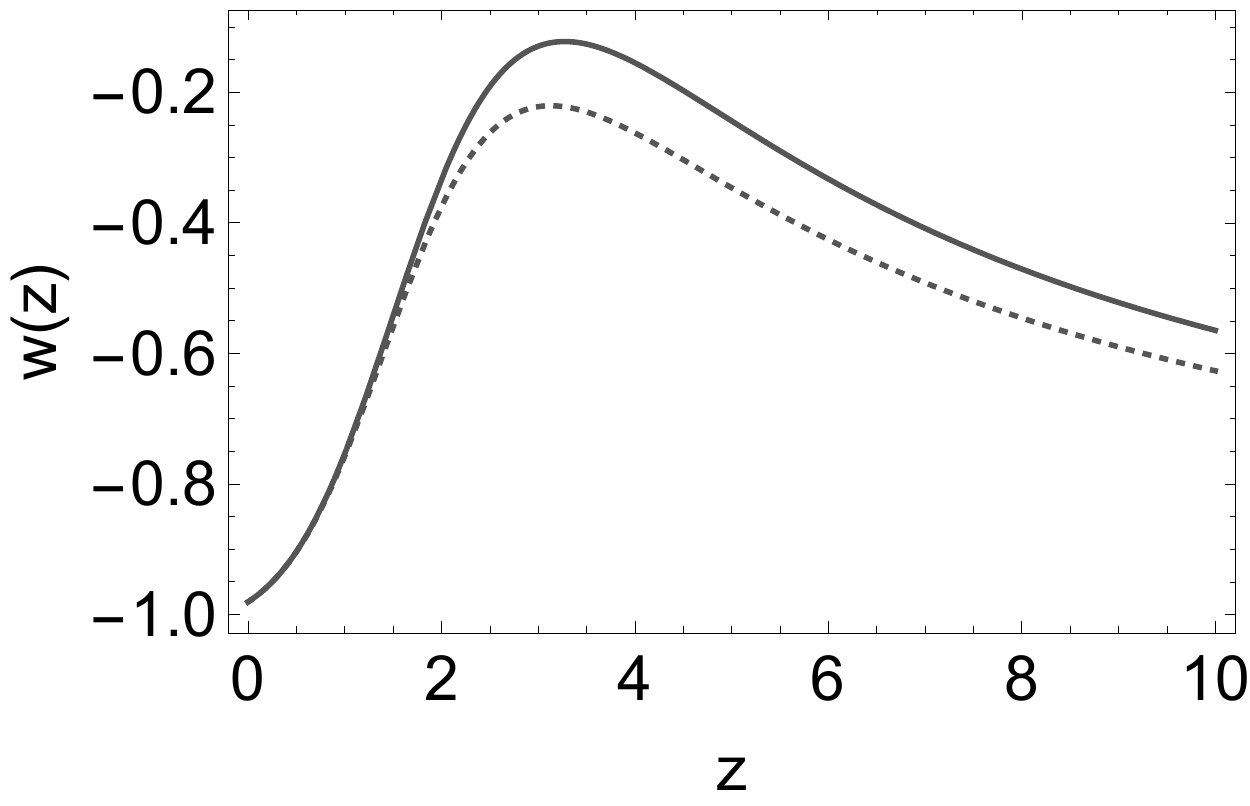}
\includegraphics[width=0.32\textwidth,origin=c,angle=0]{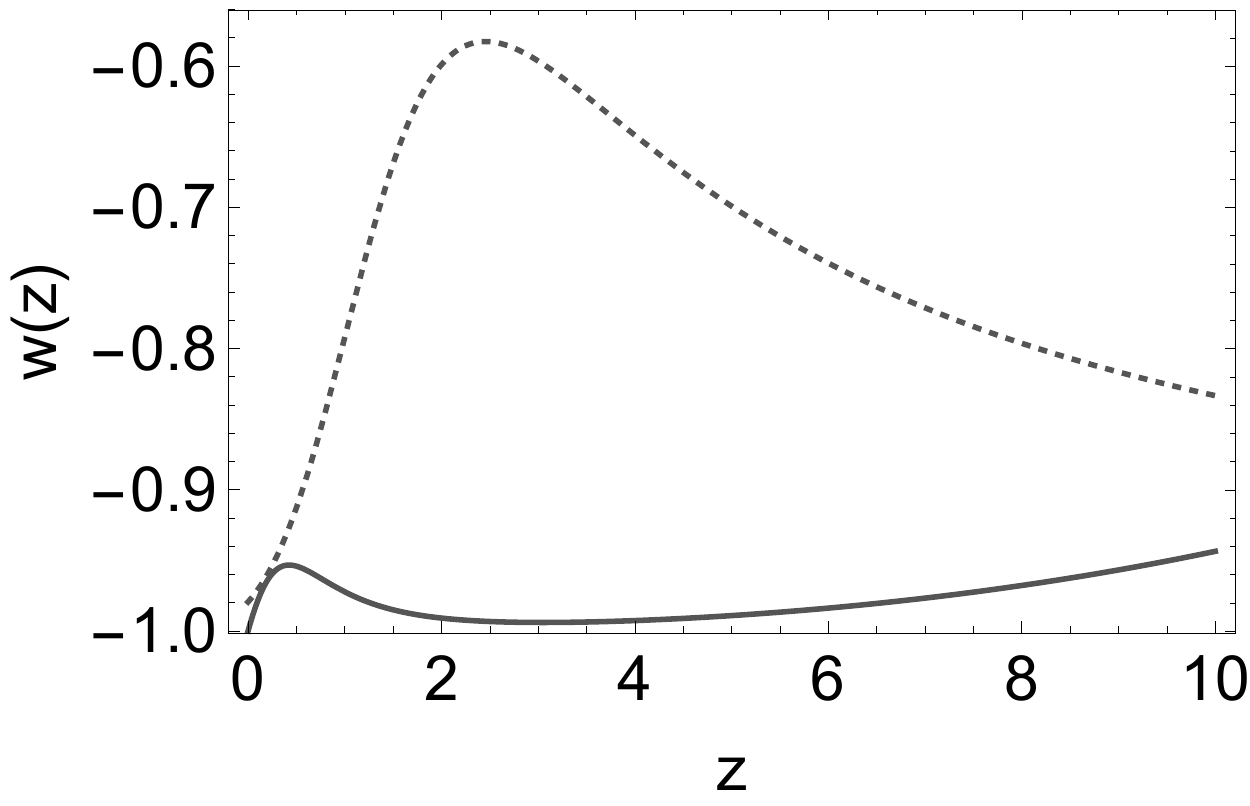}\\
\includegraphics[width=0.32\textwidth,origin=c,angle=0]{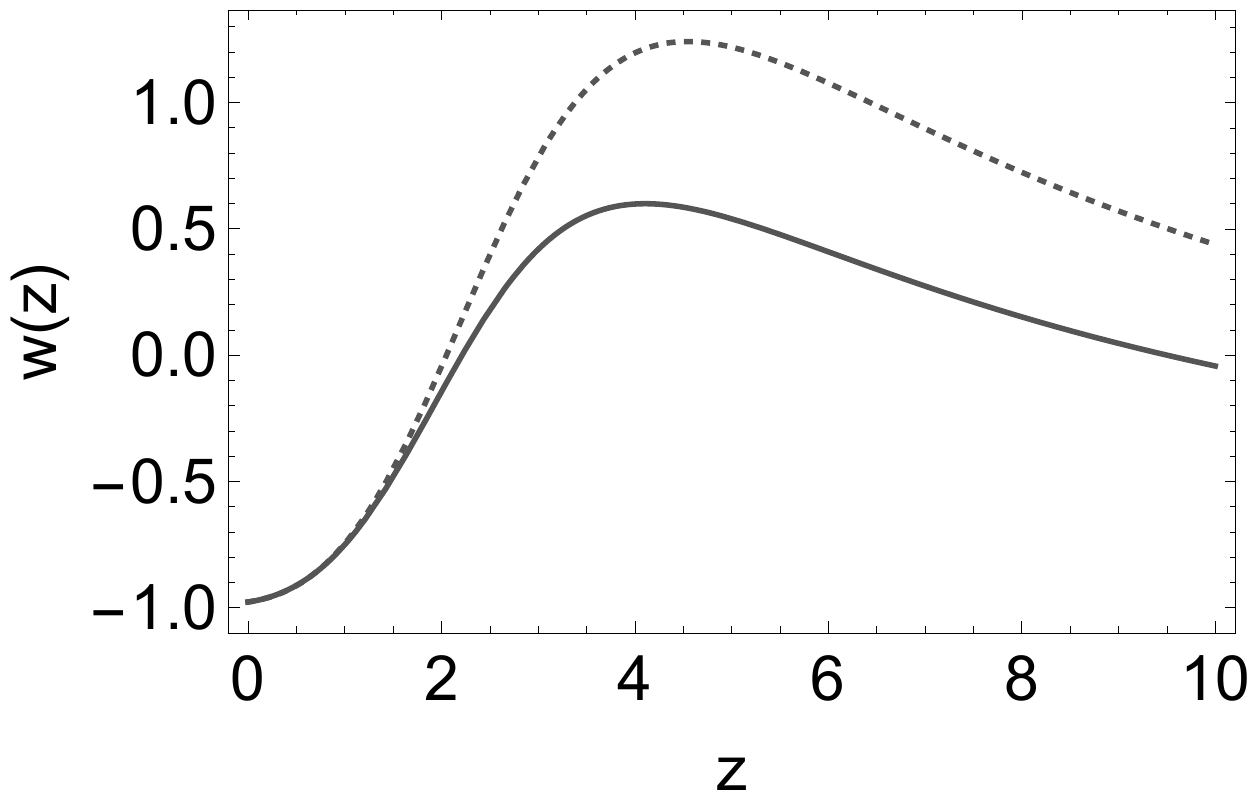}
\includegraphics[width=0.32\textwidth,origin=c,angle=0]{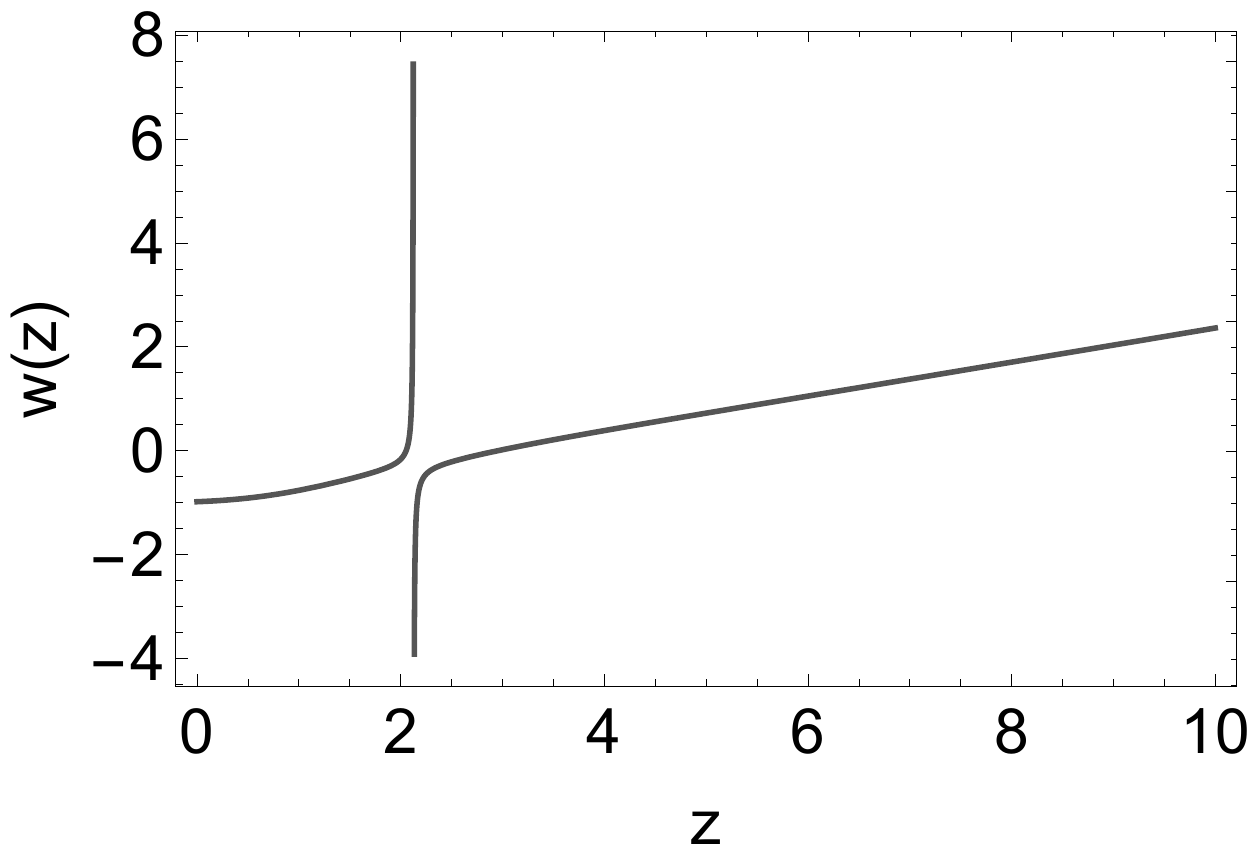}
\caption{Evolution of Power Law EoS (\ref{eq:eosz_powerlaw}). \textit{Top Left:} Case 1: solving $T$ and $B$, with $T<B$ (solid line) and $T>B$ (dashed line). \textit{Top Right:} Case 2: Varying $m$ and $k$, with $m<k$ (solid line) and $m>k$ (dashed line). \textit{Bottom Left:} Case 3: Varying $b_0$ and $t_0$, with $b_0$ negative and $f_0$ positive (solid line) and viceversa (dashed line). \textit{Bottom Right:} Case 4: varying $t_0$ and $m$ as negative values.} 
\label{evolution_powerlaw}
\end{figure}

Notice that Eq.(\ref{eq:eosz_powerlaw}) reduces to the standard $\Lambda$CDM model $w=-1$ when $f(T,B)=0$, as expected. As a first strategy, we are going to analyse the behaviour of the EoS in Eq.(\ref{eq:eosz_powerlaw}) which is associated with the $X-$fluid considering seven cases (c.f. Figure \ref{evolution_powerlaw})
\begin{itemize}
\item Case 1.1: Solving the system Eqs.(\ref{torsionscalar_frw})-(\ref{boundaryscalar_frw}) with the condition that $T<B$, i.e  we have domination of the boundary term over the torsion scalar.
\item Case 1.2: Solving the system Eqs.(\ref{torsionscalar_frw})-(\ref{boundaryscalar_frw}) with the condition that $T>B$, i.e we have domination of the torsion scalar over the boundary term)
\item Case 2.1: varying $m$ and $k$ with the condition that $m>k$.
\item Case 2.2: varying $m$ and $k$  with the condition that $k>m$.
\item Case 3.1: varying $b_0$ and $t_0$ with the condition that $b_0 < t_0$.
\item Case 3.2: varying $b_0$ and $t_0$ with the condition that $b_0 > t_0$.
\item Case 4: varying $t_0$ and $m$ as negative values.
\end{itemize}

Following the results explored, we notice that Cases 1.1 and 1.2 (cf. with Figure \ref{evolution_powerlaw} - \textit{top left}) at $z<2$ show an accelerating cosmic expansion, while after this point Case 1.1 starts to decelerate at $z=3$, meanwhile Case 1.2 still preserves this acceleration with EoS $w<-1/3$. Cases 2.1 and 2.2 (cf. with Figure \ref{evolution_powerlaw} - \textit{top right}) have an EoS with $w<-1/3$, but the latter shows an asymptotic behaviour to $\Lambda$CDM between $z=2$ and $z=4$, which then starts to grow asymptotically to the first model at large redshift. Both Cases 3.1 and 3.2 cross the phantom divided-line, below $z=2$ they are indistinguishable. Below $z=2.5$ both models can start with an EoS with $w<-1/3$, where both have an asymptotic behaviour at large redshifts which can mimic a matter phase with $w=0$  (cf. with Figure \ref{evolution_powerlaw} - \textit{bottom left}).

This is a case of an oscillating $X-$fluid EoS below $z=6$. Case 4 (cf. with Figure \ref{evolution_powerlaw} - \textit{bottom right})  has an oscillating particularity, but it experiences a divergence point due to the corresponding energy-density becoming zero.

For the model given by Eq.(\ref{eq:eosz_powerlaw}), we perform the fitted using the completed compilation of data samplers described in \S.~\ref{sec:data_analysis}.

\begin{table*}
\begin{center}
\begin{tabular}{|l|c|c|c|c|} 
 \hline 
Parameters & Best-fit & Mean$\pm\sigma$ & 95\% lower & 95\% upper \\ \hline 
$H_{0 }$ &$67.74$ & $67.74_{-1.1}^{+1.1}$ & $65.54$ & $69.89$ \\ 
$m$ &$78.93$ & $79.19_{-6.1}^{+4}$ & $70.17$ & $88.64$ \\ 
$k$ &$49.62$ & $49.81_{-1}^{+0.73}$ & $47.82$ & $51.73$ \\ 
$b_{0 }$ &$8.16e+15$ & $1.099e+16_{-1.1e+16}^{+7e+14}$ & $9.981e+11$ & $1.640e+15$ \\ 
$c_{0 }$ &$8.949e+15$ & $7.974e+15_{-6.3e+15}^{+2.8e+15}$ & $1.169e+16$ & $1.074e+16$ \\ 
\hline 
 \end{tabular} 
 \caption{Parameters and mean values for the Power law model.}
 \end{center}
\end{table*}

\begin{figure}
\centering
\includegraphics[width=0.55\textwidth,origin=c,angle=0]{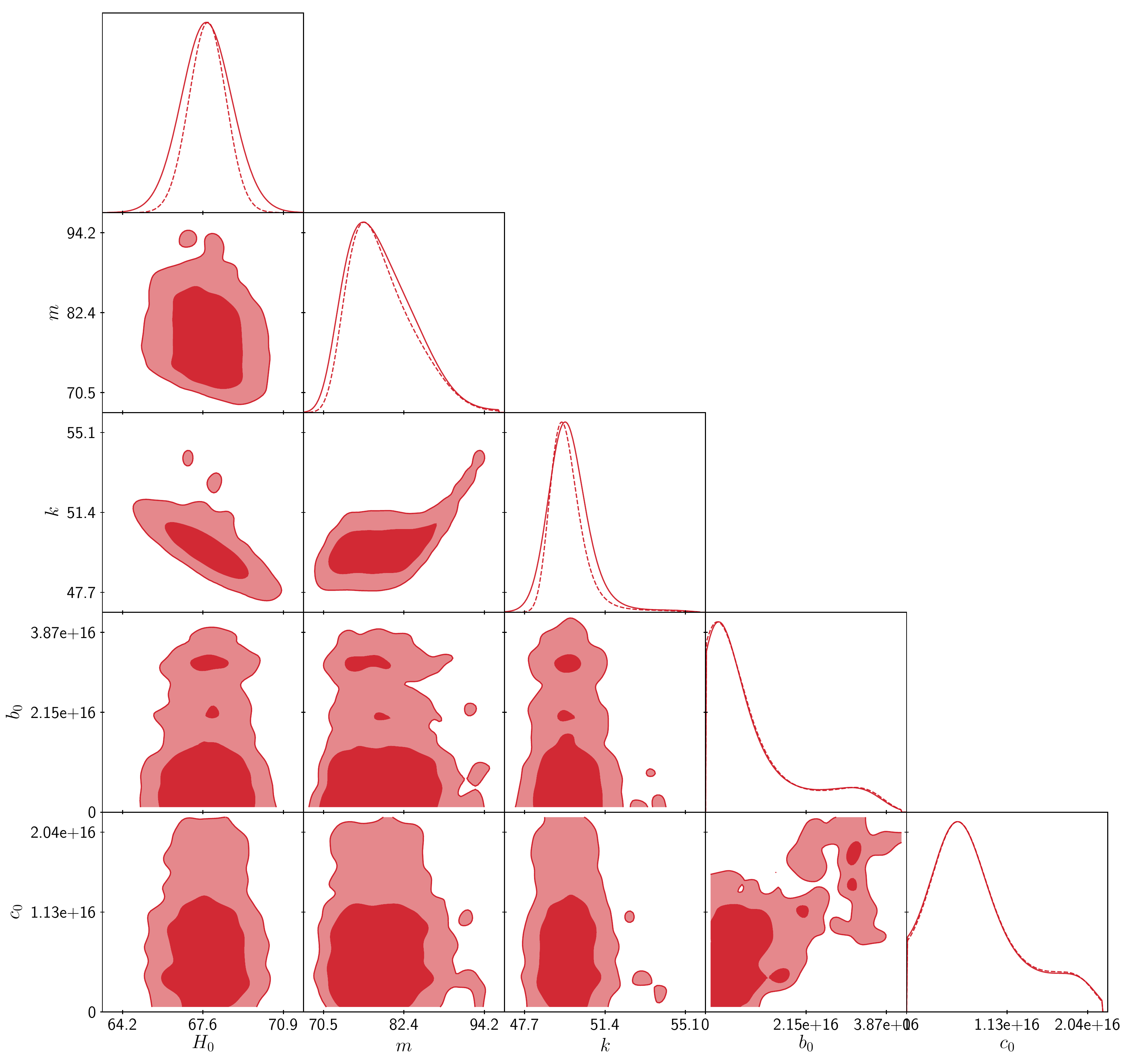}
\includegraphics[width=0.35\textwidth,origin=c,angle=0]{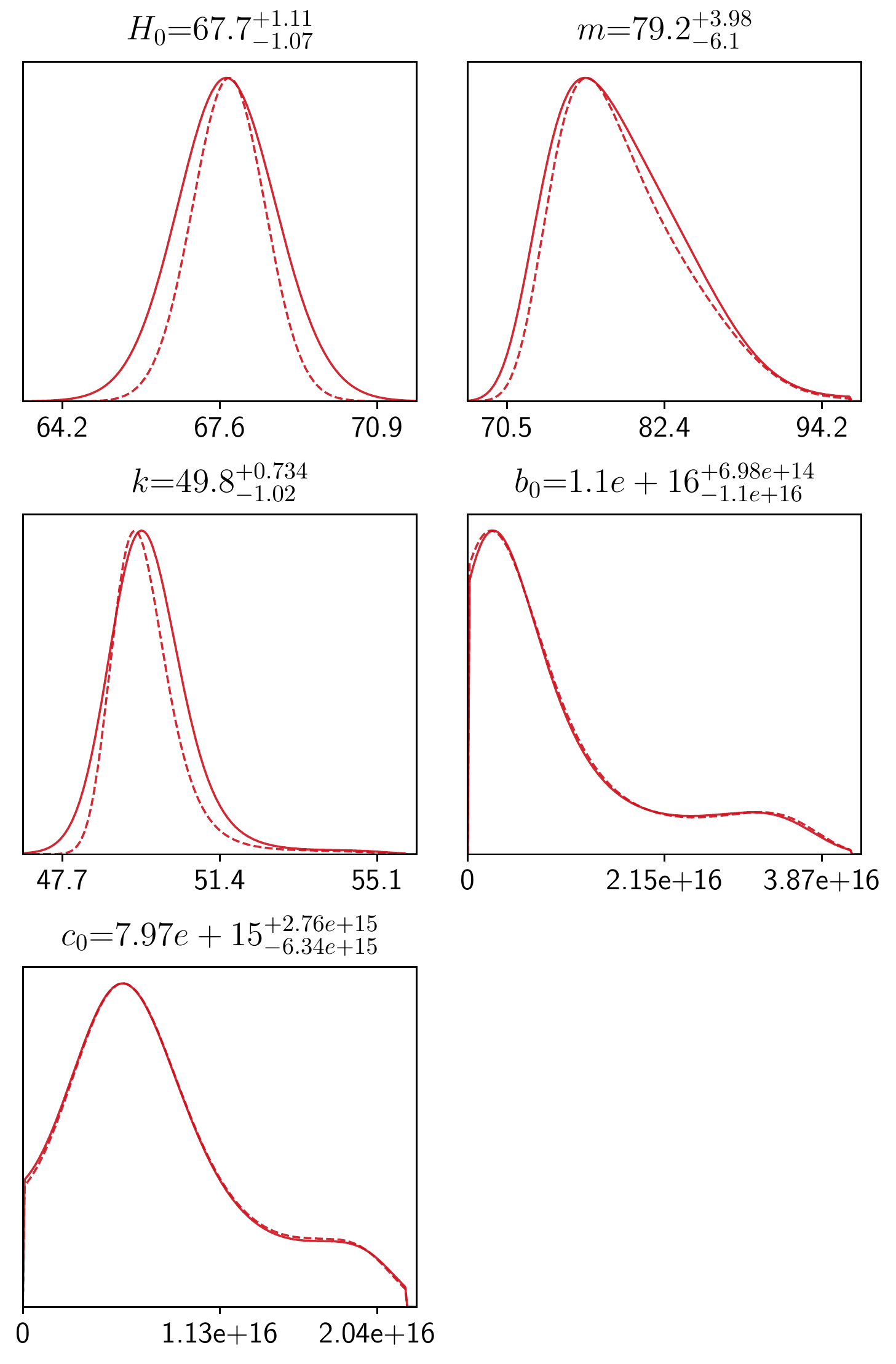}
\caption{One-dimensional marginalised distribution, and two-dimensional contours with $68\%$ and $95\%$ confidence level for the free parameters of the Power Law model using the constrained solutions for $T$ and $B$ scalars and CC+Pantheon+BAO total sampler.} 
\label{contour_powerlaw}
\end{figure}


\subsection{Mixed Power Law Model}
In Ref.\cite{Bahamonde:2016grb}, it was shown that this model can reproduce several important power law scale factors relevant for several cosmological epochs. This models takes the form
\begin{equation} \label{eq:mixed_power}
f(T,B) = f_0 B^k T^m\,,
\end{equation}
where the second- and fourth-order contributions will now be mixed, and $f_0,k,m$ are arbitrary constants. This model limits to GR when the index powers vanish, i.e. when $k=0=m$. All the derivatives corresponding to $f_T$ and $f_B$ functions for this model can be found in the Appendix \ref{app:MPL}.

As in the latter scenarios, we can compute the EoS in Eq.(\ref{EoS_func}) for the model in Eq.(\ref{eq:mixed_power}), obtaining
\begin{eqnarray}
 w_{x}(a)&&= \nonumber\\
 &&\frac{1}
   {B a(t) \left[72 \dot{a}(t)^5 w_{x_8}
   -36 \dot{a}(t)^3   a(t) w_{x_9}
   +36 a(t)^2  \dot{a}(t)^2 w_{x_{10}}
   + a(t)^5 w_{x_{11}}
   \right]}\Bigg[w_{x_1} a(t)^6\nonumber\\
   && +6 w_{x_2} a(t)^5 
  +576 w_{x_3} \dot{a}(t)^6
  +24 w_{x_4} a(t) \dot{a}(t)^4 
   +12 w_{x_5}a(t)^3 \dot{a}(t)\nonumber\\ 
   && -6 w_{x_6} a(t)^4 
   +12 w_{x_7} a(t)^2 \dot{a}(t)^2\Bigg]\,,\nonumber\\
   &&
\end{eqnarray}
where, for this case, the functions $w_{x_i}$ are given by
\begin{eqnarray}
w_{x_1}&=& -T^{m+2}f_0 B^{k+3}\,, \\
w_{x_2}&=& \left(B^2 T^2 a''(t)-2 B^k (k-1) k T^m f_0 \left(a^{(4)}(t) T^2+(m-1)a^{(3)}(t)\right)\right) B\,, \\
w_{x_3}&=& k T \left(B^k (-(k-1) (B+2 k-4)-B (k-2) m) T^m-B^3 m T\right) f_0 \,, \\
w_{x_4}&=& f_0 \left(T^m \left(4 (m-1) m B^3-(8 B+9) k m T B-18 (k-1) k T^2\right) a'(t) B^{k+1} \right. \nonumber\\&& \left.
+6 k T \left(((k-1) (7 B+12k-24)+7 B (k-2) m) T^m B^k+7 m T B^3\right) a''(t)\right)\,, \\
w_{x_5}&=& f_0 (T^m (2 B (k (m-1) (2 k+B m-2) a'(t)^2\nonumber\\
	&&+\left(-2 (m-1) m B^3+3 k m T B^2+9 (k-1) k T^2\right) a''(t) a'(t) \nonumber \\ &&
   +B^2 m (-2 m B+2 B+3 k T) a''(t)^2)+k T \Bigg(B \left(4 m B^2+9 (k-1) T\right) a'(t)\nonumber \\
    &&-12 ((k-1) (B+3 k-6) 
   +B (k-2) m) a''(t)\Bigg)a^{(3)}(t)) B^k\nonumber \\ 
   &&+2 k m T^2 a''(t) \left(5 B a'(t)-6 a^{(3)}(t)\right) B^3)\,, \\
w_{x_6}&=& 2 k T^m f_0
   \left(6 (k-2) (k-1) T a^{(3)}(t)^2-2 B T \left(m a''(t) B^2+\left(m B^2-k T+T\right) a'(t)\right) a^{(3)}(t) \right. \nonumber\\&& \left.
   +B a''(t)
   \left(3 (k-1) a''(t) T^2+(m-1) (3 k+2 B m-3) a'(t)\right)\right) B^k \nonumber\\ &&
   +T^2 \left(a'(t)^2+4 B k m f_0 a^{(3)}(t) a'(t)+4 B
   k m f_0 a''(t)^2\right) B^3\,, \\
w_{x_7}&=& f_0 \Big(B^k T^m \big(4 B \left((m-1) m B^3-2 k m T B^2-3 (k-1)
   k T^2\right) a'(t)^2 \nonumber\\
   &&+\Big[B \left(-4 (m-1) m B^3+2 (2 B+9) k m T B+27 (k-1) k T^2\right) a''(t)\nonumber\\
   &&+12 k ((k-1) (B+4
   k-8)+B (k-2) m) T a^{(3)}(t)\big) \nonumber \\ 
   && \times   a'(t)-18 k ((k-1) (2 B+3 k-6)+2 B (k-2) m) T a''(t)^2\Big)\nonumber\\
   &&-6 B^3 k m T^2 \left(B
   a'(t)^2-2 a^{(3)}(t) a'(t)+6 a''(t)^2\right)\Big]\,,
   \end{eqnarray}
   \begin{eqnarray}
   w_{x_8}&=& T^m \left(-2 (m-1) m B^3+(4 B+3) k m T B+6 (k-1) kT^2\right) f_0 B^k\,, \\
   w_{x_9}&=& T^m f_0 \Big(2 k T (B m+2 (k-1) T) a'(t)\nonumber\\
   &&+\left(-4 (m-1) m B^3+6 (B+1) k m TB+9 (k-1) k T^2\right) a''(t)\Big) B^k\,, \\
%
   w_{x_{10}}&=&k T^{m+1} f_0  \left((2 B m+3 (k-1) T) a''(t)+\left(-2 mB^2-3 k T+3 T\right) a^{(3)}(t)\right) B^k\,, \\
   w_{x_{11}}&=& T^{m+2} f_0 B^{k+2}-6 T^2 a(t)^4 a''(t) B^2+6 T^2 a(t)^3 a'(t)\left(6 (k-1) k T^m f_0 a^{(3)}(t) B^k+a'(t) B^2\right)\,,\nonumber\\
   &&
\end{eqnarray}

Again, we rewrite the above expression in terms of the redshift $z=a_0 /a -1$, therefore the mixed power law model EoS can be expressed as
\begin{equation}\label{eq:eosz_mixedpowerlaw}
w(z) = \frac{(z+1) \left[w(z)_1+w(z)_2 +24 w(z)_3
   +12w(z)_4
   -12w(z)_5
   +576 w(z)_6
   -6 w(z)_7
   \right]}
   {B \left[w(z)_8
   +36 w(z)_9
   -w(z)_{10}
   +w(z)_{11}\right]}\,,
\end{equation}
where
\begin{eqnarray}
w(z)_1 &=&-\frac{T^{m+2} f_0 B^{k+3}}{(z+1)^6}\,, \\
w(z)_2 &=& \frac{6 \left(\frac{2 B^2 T^2}{(z+1)^3}-2 B^k (k-1) k T^m
   \left(\frac{24 T^2}{(z+1)^5}-\frac{6 (m-1)}{(z+1)^4}\right) f_0\right) B}{(z+1)^5}\,, \\
   (z+1)^9 f_0 w(z)_3 &=& \frac{12 k T
   \left(((k-1) (7 B+12 k-24)+7 B (k-2) m) T^m B^k+7 m T B^3\right)}{(z+1)^3} \nonumber \\&&
   -\frac{B^{k+1} T^m \left(4 (m-1) m B^3-(8
   B+9) k m T B-18 (k-1) k T^2\right)}{(z+1)^2} \,, \\
%
  (z+1)^6  w(z)_4&=& B^k T^m \left(-\frac{72 k ((k-1) (2
   B+3 k-6)+2 B (k-2) m) T}{(z+1)^6}- w(z)_{12}   +w(z)_{13}\right) \nonumber \\&&
   -6 B^3 k m T^2 \left(\frac{B}{(z+1)^4}+\frac{12}{(z+1)^6}\right)
   f_0\,, \\
   w(z)_5&=& \frac{T^m w(z)_{14} B^k+\frac{4 k m T^2 \left(\frac{36}{(z+1)^4}-\frac{5 B}{(z+1)^2}\right)
   B^3}{(z+1)^3} f_0}{(z+1)^5}\,, \\
   w(z)_6 &=&  \frac{k T \left(B^k ((1-k) (B+2 k-4)-B (k-2) m) T^m-B^3 m T\right)
   f_0}{(z+1)^{12}}\,, \\
   (z+1)^4 w(z)_7   &=&2k T^m\nonumber \\ 
   && \left(w(z)_{15}+\frac{216 (k-2) (k-1) T}{(z+1)^8}+\frac{2 B \left[\frac{6 (k-1)
   T^2}{(z+1)^3}-\frac{(m-1) (3 k+2 B m-3)}{(z+1)^2}\right]}{(z+1)^3}\right) f_0 B^k \nonumber\\
   &&   +T^2 \left(\frac{40 B k m
   f_0}{(z+1)^6}+\frac{1}{(z+1)^4}\right) B^3\,,
   \end{eqnarray}
  
   \begin{eqnarray}
   w(z)_8&=& \frac{36 k T^{m+1} \left[\frac{2 (2 B m+3
   (k-1) T)}{(z+1)^3}-\frac{6 \left(-2 m B^2-3 k T+3 T\right)}{(z+1)^4}\right] f_0 B^k}{(z+1)^6}\,, \\
   w(z)_9 &=& \frac{T^m
   \left[\frac{2 \left(-4 (m-1) m B^3+6 (B+1) k m T B+9 (k-1) k T^2\right)}{(z+1)^3}-\frac{2 k T (B m+2 (k-1)
   T)}{(z+1)^2}\right] f_0 B^k}{(z+1)^7}\,, \\
   w(z)_{10}&=& \frac{72 T^m \left(-2 (m-1) m B^3+(4 B+3) k m T B+6 (k-1) k T^2\right) f_0
   B^k}{(z+1)^{10}}\,, \\
   w(z)_{11}&=& \frac{T^{m+2} f_0 B^{k+2}}{(z+1)^5}-\frac{12 T^2 B^2}{(z+1)^7}-\frac{6 T^2 \left(-\frac{36 (k-1) k T^m
   f_0 B^k}{(z+1)^4}-\frac{B^2}{(z+1)^2}\right)}{(z+1)^5}\,, \\
   w(z)_{12}&=& \frac{\frac{2 B \left(-4 (m-1) m B^3+2 (2 B+9) k m T B+27 (k-1) k
   T^2\right)}{(z+1)^3}-\frac{72 k ((k-1) (B+4 k-8)+B (k-2) m) T}{(z+1)^4}}{(z+1)^2}\,, \nonumber\\
   &&\\
   w(z)_{13}&=& \frac{4 B \left[(m-1) m B^3-2 k m T
   B^2-3 (k-1) k T^2\right]}{(z+1)^4}\,,
    \end{eqnarray}
   
    \begin{eqnarray}
   w(z)_{14}&=& 2 B \left( \frac{4 m (-2 m B+2 B+3 k T) B^2}{(z+1)^6}+\frac{k (m-1) (2 k+B
   m-2)}{(z+1)^4} \right. \nonumber \\ && \left.
   -\frac{2 \left(-2 (m-1) m B^3+3 k m T B^2+9 (k-1) k T^2\right)}{(z+1)^5} \right) \nonumber \\ &&
   -\frac{6 k T\left(-\frac{24 ((k-1) (B+3 k-6)+B (k-2) m)}{(z+1)^3}-\frac{B \left(4 m B^2+9 (k-1)
   T\right)}{(z+1)^2}\right)}{(z+1)^4}\,, \\
   w(z)_{15}&=&\frac{12 B \left[\frac{2 B^2 m}{(z+1)^3}-\frac{m B^2-k
   T+T}{(z+1)^2}\right] T}{(z+1)^4}\,,
\end{eqnarray}

\begin{figure}
\centering
\includegraphics[width=0.35\textwidth,origin=c,angle=0]{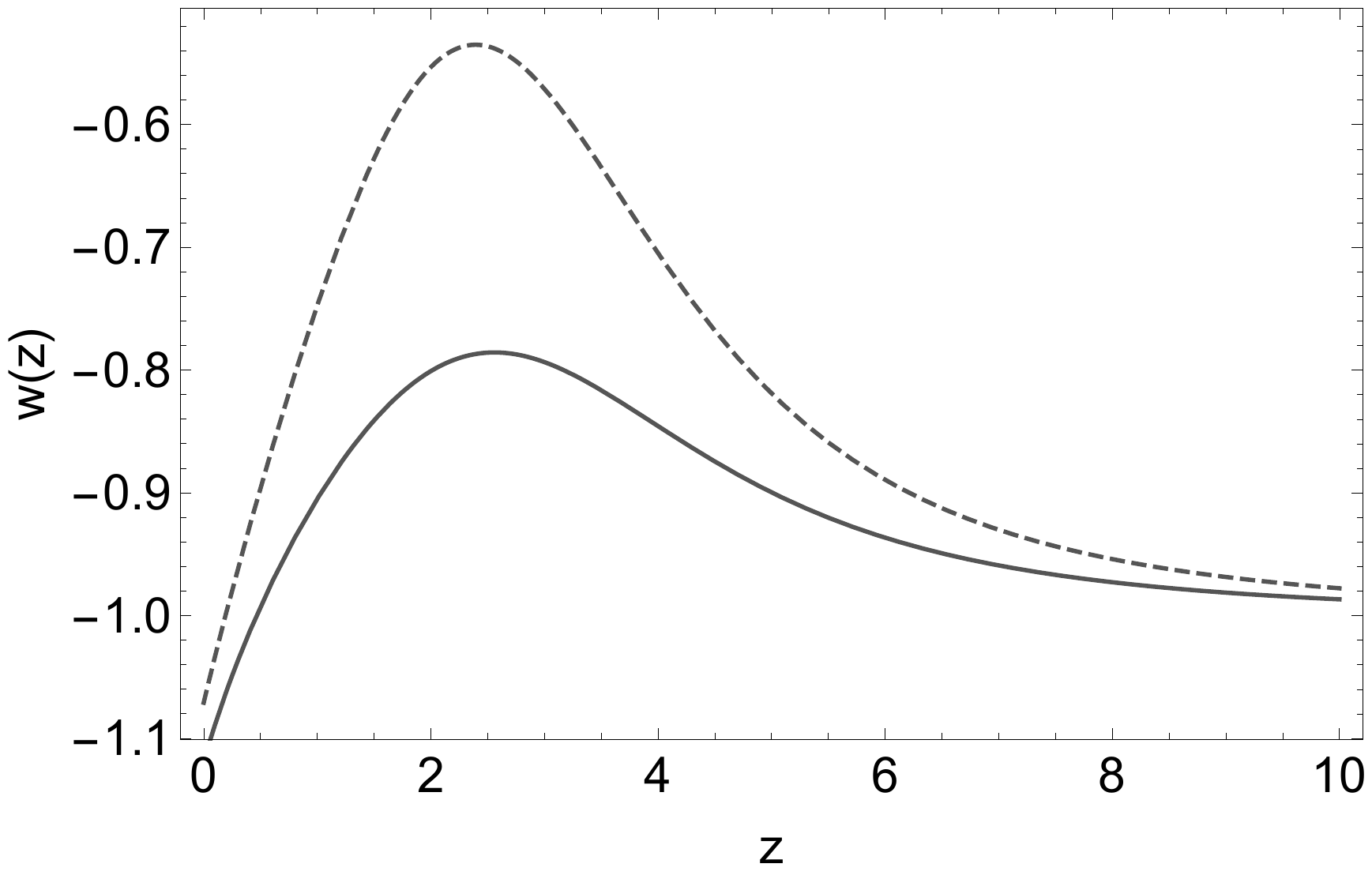}
\includegraphics[width=0.35\textwidth,origin=c,angle=0]{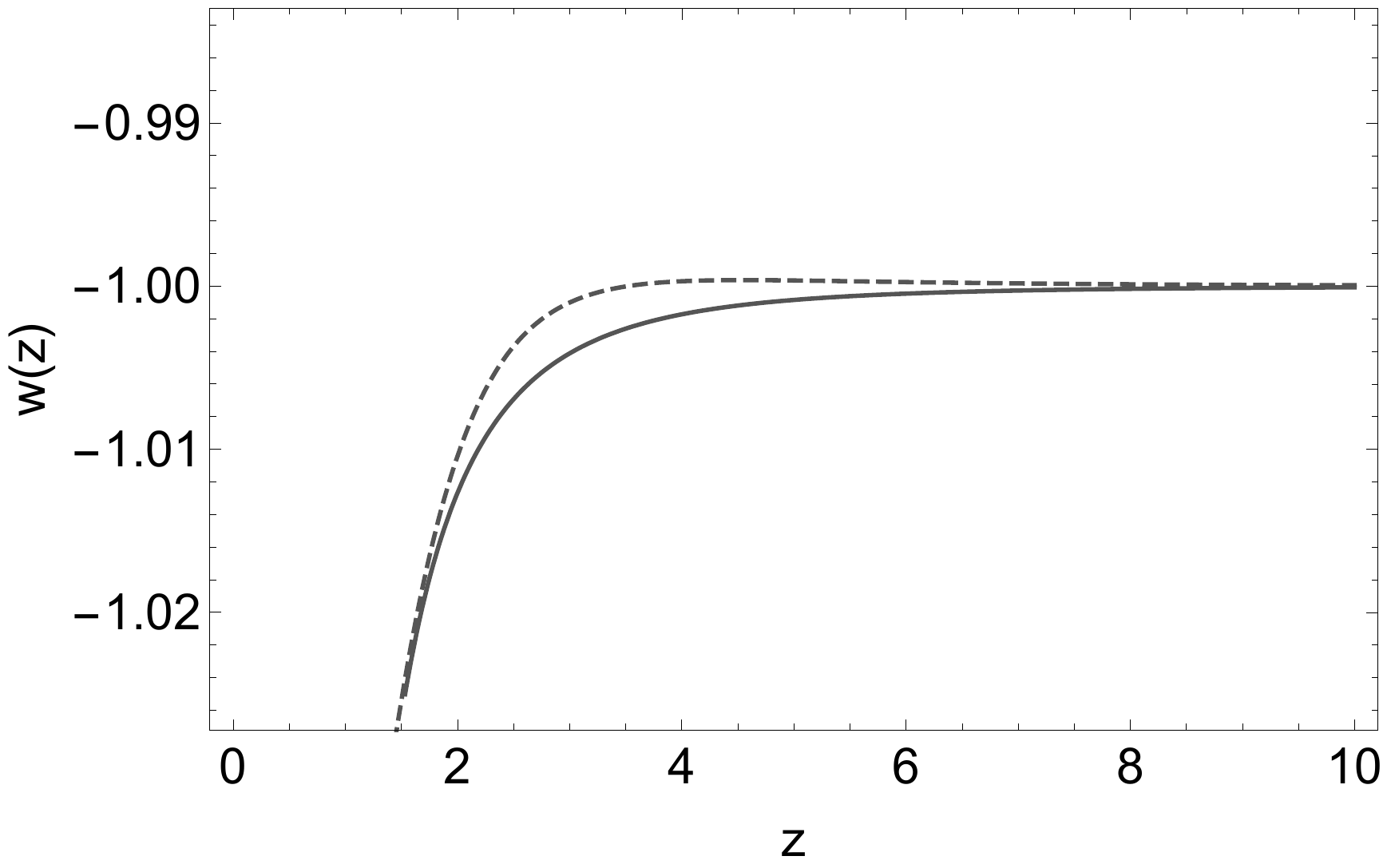}
\caption{Evolution of Mixed Power Law EoS (\ref{eq:eosz_mixedpowerlaw}). \textit{Left:} Case 1: resolving $T$ and $B$, with $T<B$ (solid line) and $T>B$ (dashed line). \textit{Right:} Case 2: Varying $m$ and $k$, with $m<k$ (solid line) and $m>k$ (dashed line).} 
\label{evolution_mixedpower}
\end{figure}

Analysing the behaviour of the EoS in Eq.(\ref{eq:eosz_mixedpowerlaw}) which is associated with the $X-$fluid considering X cases (c.f. Figure \ref{evolution_mixedpower}) and with $f_0$ as a positive constant
\begin{itemize}
\item Case 1.1: Resolving the system Eqs.(\ref{torsionscalar_frw})-(\ref{boundaryscalar_frw}) with the condition that $T<B$, i.e  we have domination of the boundary term over the torsion scalar.
\item Case 1.2: Resolving the system Eqs.(\ref{torsionscalar_frw})-(\ref{boundaryscalar_frw}) with the condition that $T>B$, i.e we have domination of the torsion scalar over the boundary term)
\item Case 2.1: varying $m$ and $k$ with the condition that $m>k$.
\item Case 2.2: varying $m$ and $k$  with the condition that $k>m$.
\end{itemize}

As in Cases 1 and 2 of the Power Law model, the Cases 1.1 and 1.2  (cf. with Figure \ref{evolution_mixedpower} - \textit{left}) in this model cross the phantom divided-line but preserve its quintessence behaviour until at high redshift both scenarios limiting to $\Lambda$CDM model. For the Cases 2 (cf. with Figure \ref{evolution_mixedpower} - \textit{right}) , at $z<4$ both scenarios mimic a phantom energy.  

For the model given by Eq.(\ref{eq:eosz_mixedpowerlaw}), we perform the fitted using the completed compilation of data samplers described in \S.~\ref{sec:data_analysis}.

\begin{table*}
\begin{center}
\begin{tabular}{|l|c|c|c|c|} 
 \hline 
Parameter & best-fit & mean$\pm\sigma$ & 95\% lower & 95\% upper \\ \hline 
$H_{0 }$ &$67.92$ & $67.86_{-1.1}^{+1.2}$ & $65.63$ & $70.1$ \\ 
$m$ &$36.59$ & $38.02_{-2.6}^{+2.1}$ & $33.71$ & $42.45$ \\ 
$k$ &$2.55$ & $2.626_{-0.13}^{+0.11}$ & $2.396$ & $2.861$ \\ 
$c_{0 }$ &$1.653e+11$ & $4.97e+11_{-5e+11}^{+-2.6e+11}$ & $2.353e+09$ & $4.847e+09$ \\ 
\hline 
 \end{tabular} \caption{Parameters and mean values for the Mixed Power Law Model model.}
 \end{center}
\end{table*}

\begin{figure}
\centering
\includegraphics[width=0.5\textwidth,origin=c,angle=0]{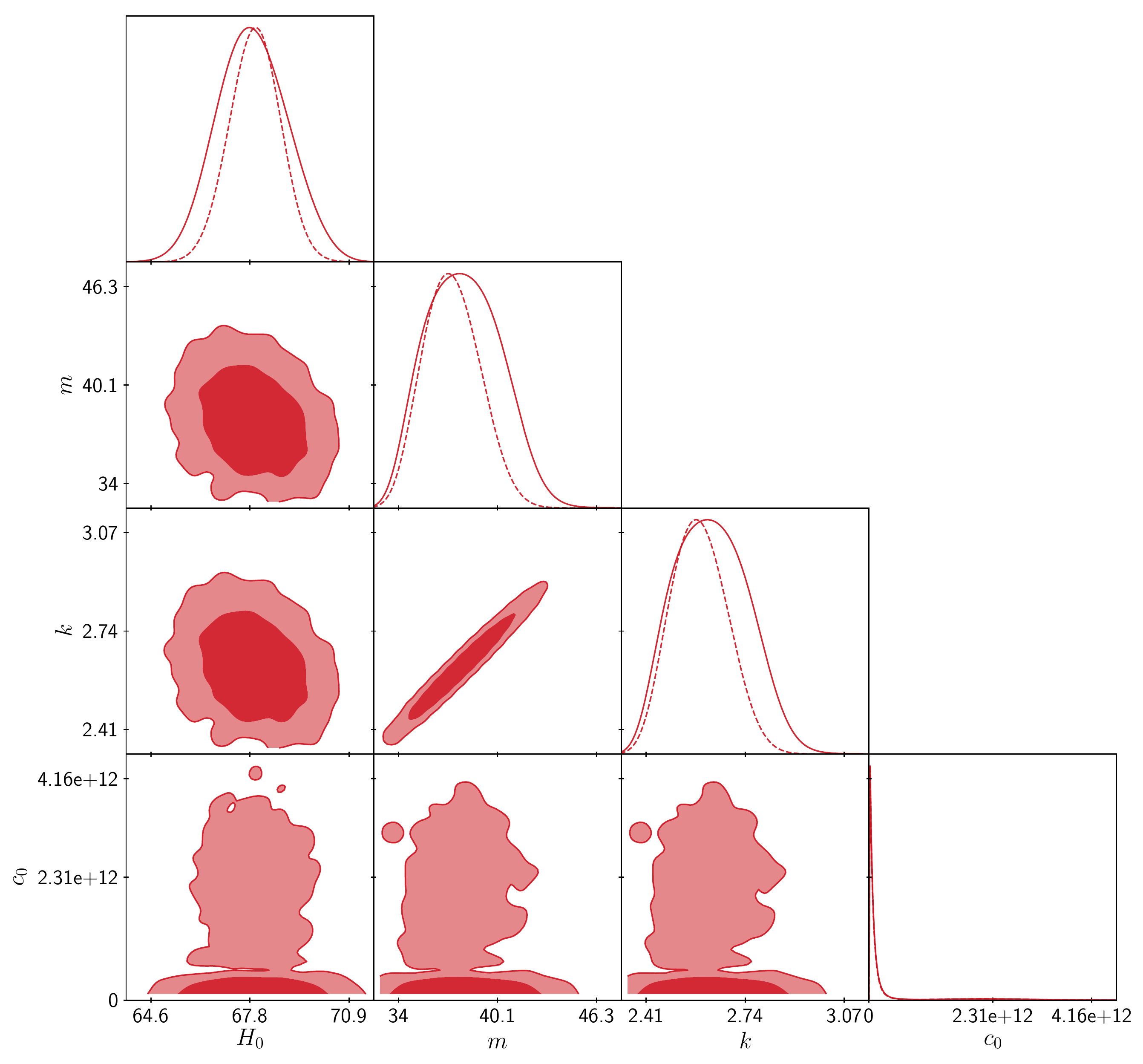}
\includegraphics[width=0.4\textwidth,origin=c,angle=0]{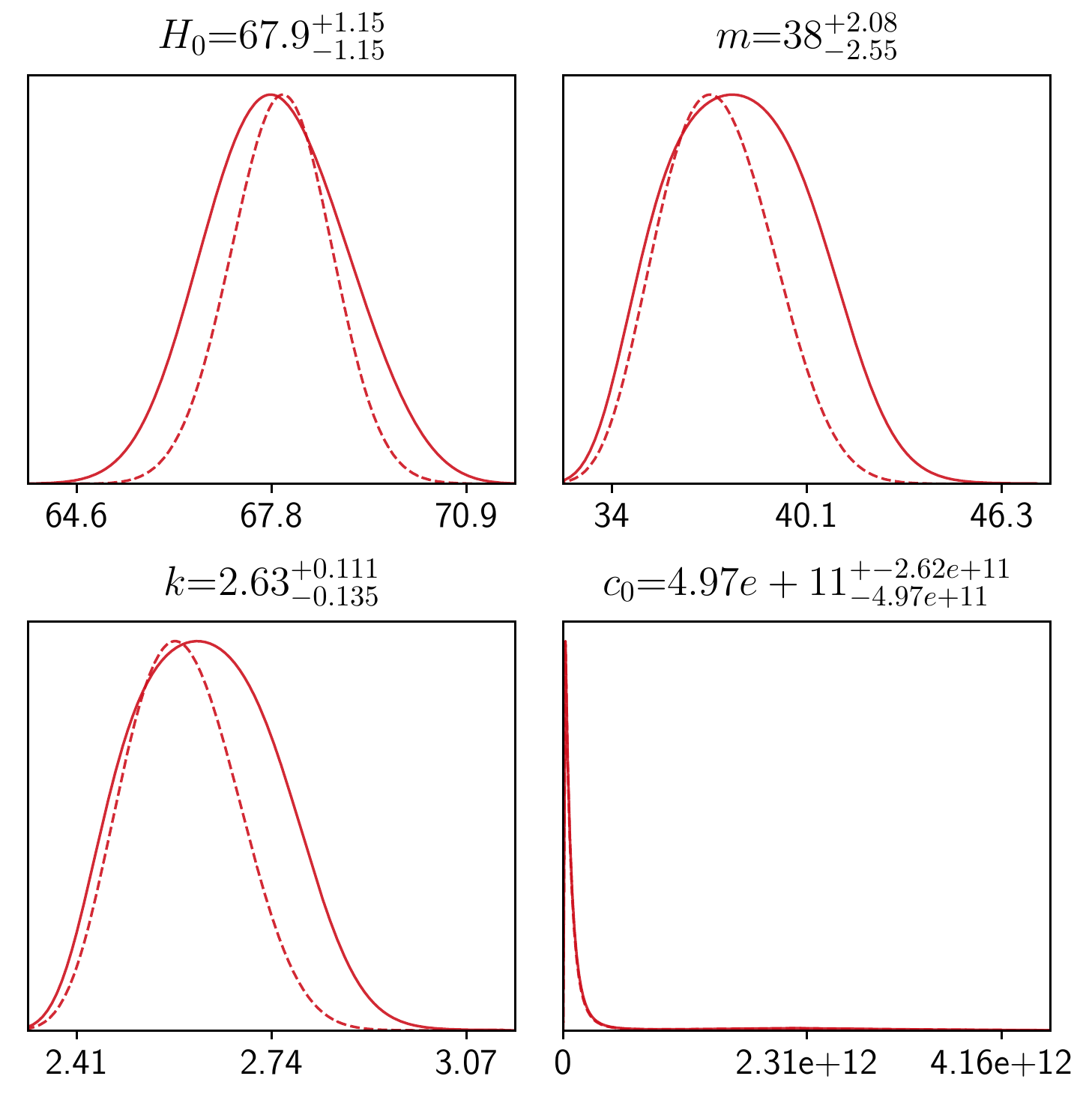}
\caption{One-dimensional marginalised distribution, and two-dimensional contours with $68\%$ and $95\%$ confidence level for the free parameters for the Mixed Power Law model using the constrained solutions for $T$ and $B$ scalars and CC+Pantheon+BAO total sampler.} 
\label{contour_mixedlaw}
\end{figure}


\subsection{Boundary Term Deviations to TEGR model}
In general, we can embody modifications to TEGR as
\begin{equation} \label{eq:TEGR}
f(T,B) = -T + g(B),
\end{equation}
where modifications to standard gravity are chosen to be expressed through contributions from the boundary term only. One interesting model also investigated in Ref.\cite{Bahamonde:2016grb} is the one where
\begin{equation} \label{eq:form_B}
g(B) = f_1 B\ln B\,,
\end{equation}
where $f_1$ is an arbitrary constant. 

\begin{figure}
\centering
\includegraphics[width=0.4\textwidth,origin=c,angle=0]{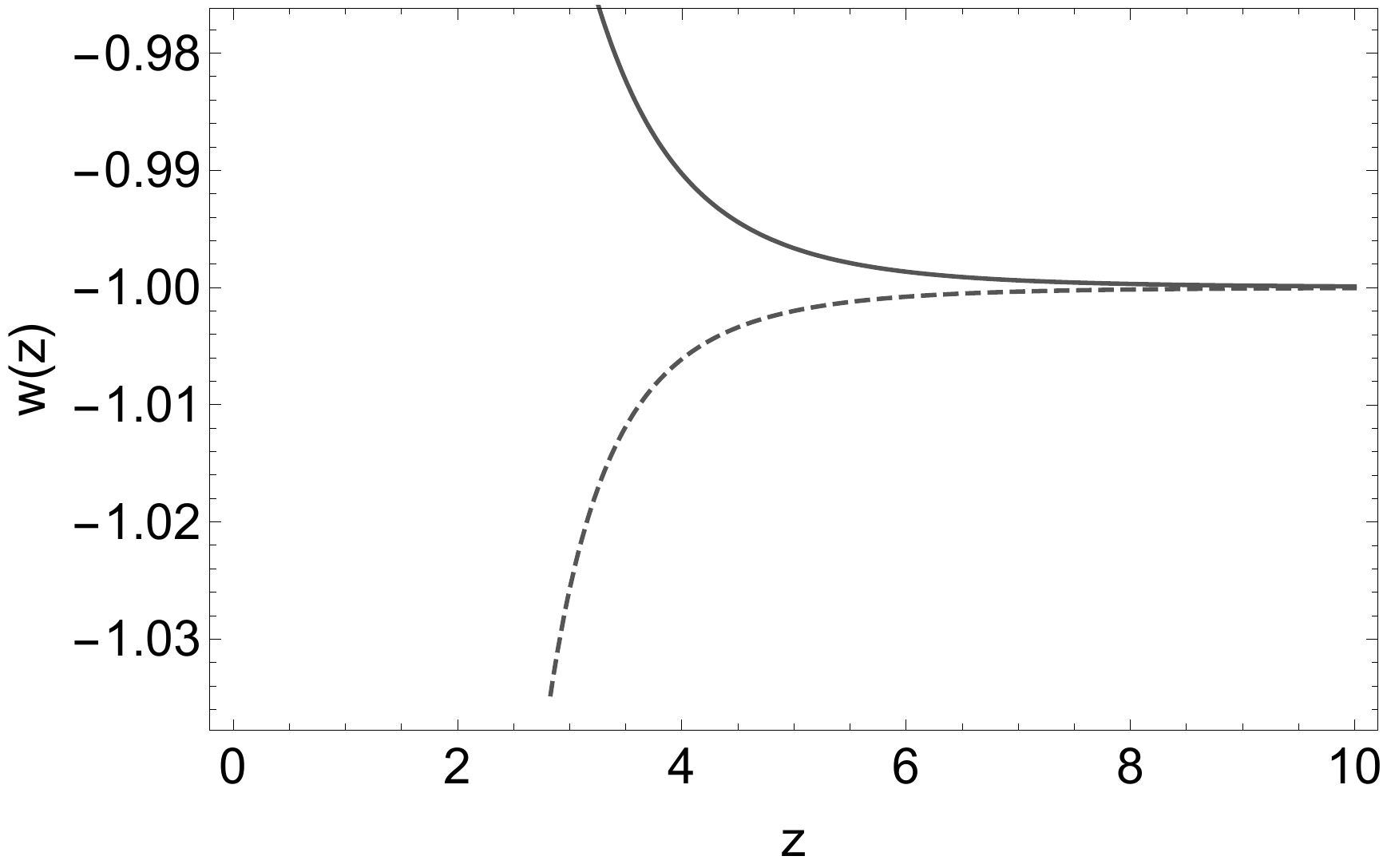}
\caption{Evolution of TEGR EoS (\ref{eq:eosz_TEGR}). Here we solved $T$ and $B$, with $T<B$ (solid line) and $T>B$ (dashed line).} 
\label{evolution_TEGR}
\end{figure}

As in the latter scenarios, we can compute the EoS (\ref{EoS_func}) for (\ref{eq:TEGR}) and obtain
\begin{eqnarray}
 w_{x}(a)&=& \frac{1152 f_1 \dot{a}(t)^6+6 B a(t)^5 w_{x_1} 
+ w_{x_8}
   -36 f_1
   a(t)^2 \dot{a}(t)^2 w_{x_5}
   +B^2 a(t)^6 \left(T-B f_1 \ln B\right)}
   {B a(t) \left[w_{x_6}
   -36 w_{x_7}+B a(t)^5 \left(B f_1 \ln B-T\right)\right]}\,,\nonumber\\
   &&
\end{eqnarray}
where
\begin{eqnarray}
w_{x_1} &=& B \ddot{a}(t)-2 f_1 a^{(4)}(t)\,, \\
w_{x_2}&=& 4 \ddot{a}(t)+B\dot{a}(t)\,, \\
w_{x_3}&=& B^2 \dot{a}(t)^2+6 f_1 \left[B \ddot{a}(t)^2-2 a^{(3)}(t)^2\right]+4 B f_1 a^{(3)}(t)\,, \\
w_{x_4} &=& 4 a^{(3)}(t) \ddot{a}(t)+B a'(t) \left[a^{(3)}(t)+2 \ddot{a}(t)\right]\,, \\
w_{x_5} &=& -18 \ddot{a}(t)^2+4 B \dot{a}(t)^2+ \dot{a}(t) \left[16 a^{(3)}(t)-9 B \ddot{a}(t)\right]\,, \\
w_{x_6}&=& -6 B a(t)^4 \ddot{a}(t)+432 f_1 \dot{a}(t)^5+6 a(t)^3 \dot{a}(t) \left[6 f_1 a^{(3)}(t)+B\dot{a}(t)\right]\,, \\
w_{x_7}&=& f_1 a(t) \dot{a}(t)^3 \left[9 \ddot{a}(t)+4 \dot{a}(t)\right]+108 f_1 a(t)^2 \dot{a}(t)^2
   \left[\ddot{a}(t)-a^{(3)}(t)\right], \\
   w_{x_8}&=&  -432 f_1 a(t) \dot{a}(t)^4 w_{x_2}
   -6 a(t)^4 w_{x_3}
   +108 f_1 a(t)^3 \dot{a}(t) w_{x_4}\,,
\end{eqnarray}
We rewrite the above expression in terms of the redshift $z=a_0 /a -1$, which leads to the mixed power law model EoS being expressed as
\begin{eqnarray}\label{eq:eosz_TEGR}
w(z)&=&\frac{B^2 (z+1)^4 \left[T (z+1)^2+6\right]-f_1 \left[B^3 (z+1)^6 \ln B+288 (B (z+1) (z+4)-16)\right]}{B (z+1)
   \left[f_1 \left(B^2 (z+1)^5 \ln B+288 (z+4)\right)-B (z+1)^3 \left(T (z+1)^2+6\right)\right]}\,.\nonumber\\
   &&
\end{eqnarray}
As we can see from Eq.(\ref{eq:form_B}), we need the existence of the boundary scalar with a constraint equation given by $3H^2 +\dot{H}=1/6$ (for N=1) and $f_{1}>0.1$ in order to avoid singularities on the EoS in Eq.(\ref{eq:eosz_TEGR}). The behaviour for this case is show in Figure \ref{evolution_TEGR}, where we notice that when the boundary scalar dominates over the torsion scalar its EoS mimics a quintessence fluid, while on the contrary, we notice solely a phantom behaviour. Both scenarios limiting to a $\Lambda$CDM model at larger redshifts.

For the model given by (\ref{eq:eosz_TEGR}), we perform the fitted using the completed compilation of data samplers described in \S.~\ref{sec:data_analysis}.

\begin{table*}
\begin{center}
\begin{tabular}{|l|c|c|c|c|} 
 \hline 
Parameter & best-fit & mean$\pm\sigma$ & 95\% lower & 95\% upper \\ \hline 
$H_{0 }$ &$79.77$ & $79.83_{-0.91}^{+0.9}$ & $78.03$ & $81.64$ \\ 
$f_{1 }$ &$0.203$ & $0.203_{-0.002}^{+0.002}$ & $0.199$ & $0.208$ \\ 
\hline 
 \end{tabular} 
  \caption{Parameters and mean values for the TEGR Deviations model.}
  \end{center}
\end{table*}

\section{Conclusions}
\label{sec:conclusions}
In this paper we presented cosmological analyses for $f(T,B)$ theory with a flat homogeneous and isotropic metric. Also, we obtained a generic equation of state $w_{\mbox{eff}}$ in Eq.(\ref{EoS_func}) from which we can consider any cosmological form for the torsion scalar and boundary terms. As a first approach, we proposed four $f(T,B)$ cosmological scenarios where we obtained:
\begin{enumerate}
\item The analytical solutions for the $w(z)$ for each model and,
\item full data analyses, where we were capable to constrain the free cosmological parameters for our models using a total sampler of CC+PantheonSNeIa+BAO, fitted with Planck 2018 and SH0ES+H0LiCOW, respectively.
\end{enumerate}

The main results for these cosmological models are:
\begin{itemize}
\item General Taylor Expansion Model: According to Eq.(\ref{eq:eosz_taylor}), where after solving the system of equations in Eqs.(\ref{torsionscalar_frw})-(\ref{boundaryscalar_frw}), we divided the analytical solutions in several cases: from 1.1 to 2.2. For each case it is possible to see the domination of the torsion or boundary terms. Notice that the $A_{i}$ parameters need to be positive, if not we have singularities. 
\item Power Law Model: As in the later case, we obtained analytical solutions for Eq.(\ref{eq:eosz_powerlaw}) and divided in cases 1.2 till 4. On one hand, we notice the equation of state for this scenario reduces to the standard $\Lambda$CDM model $w=-1$ when $f(T,B)=0$. On the other hand,
this is an interesting scenario were we have an oscillating $w(z)$, but when observational samplers are used, this model wants to remain phantom-like. This scenario seems to be in agreement with the value $H_0$ given by Planck 2018 with a difference of $0.3$-$\sigma$, c.f. with Figure \ref{contour_powerlaw}.
\item Mixed Power Law Model: For this scenario we have Cases 1.1 till 2.2. It can reproduce $\Lambda$CDM at high redshifts. This scenario, as the latter, seems to be in agreement with the value of $H_0$ given by Planck 2018 with a difference of $0.5$-$\sigma$, but $c_0$ is quite correlated with $H_0$, c.f. with Figure \ref{contour_mixedlaw}.
\item Boundary Term Deviations to TEGR model: We can recover again $\Lambda$CDM for two cases. The value of the free parameter $f_1$ seems to remain small with the entire sampler used. Even more, there is an intriguing result: the $H_0$ value is greater than the reported by SH0ES+H0LiCOW, even higher than the one using SBF calibrations \cite{Verde:2019ivm}.  
\end{itemize}

On the practical level, $f(T,B)$ acts as a generalization of $f(\mathring{R})$ gravity in which the torsion scalar and boundary term contributions are decoupled from each other. For a flat FLRW cosmology, this means that some of the observationally interesting $f(\mathring{R})$ models may allow for more freedom. For instance in Ref.\cite{Nunes:2016drj}, the authors probe four popular models against CC, BAO and SNeIA data with some favouring nonzero model parameters. In Ref.\cite{delaCruz-Dombriz:2015tye} some of these models are also studied with the parameter space of $\Lambda$CDM being largely sustained despite the model modification. In other works \cite{Lazkoz:2018aqk,2010JCAP...11..004G}, the situation remains unclear whether model deviations from $\Lambda$CDM are favoured observationally. It would be interesting to use the tools of $f(T,B)$ gravity to probe potential additions from the decoupled scalars to further approximate different phenomena in cosmology.

All above results lead us to believe that an extension in the form of $f(T,B)$ gravity could produce interesting scenarios, where we can consider the role of torsion and boundary terms, fitted with astrophysical data, shedding light in the study of the late-time accelerating Universe.
According to our data analyses, this direction could be an accurate expansion of the standard cosmological model to merit further research in order to pursuit a fit more precise from widely different cosmic epochs, e.g for early universe, which will be a study that we will report in future work.


\section{Acknowledgments}\label{sec:acknowledgements} 
CE-R is supported by the Royal Astronomical Society as FRAS 10147, PAPIIT Project IA100220 and ICN-UNAM projects. The authors thankfully acknowledge computer resources and support provided by C. Nahmad. This article is based upon work from CANTATA COST (European Cooperation in Science and Technology) action CA15117, EU Framework Programme Horizon 2020. The authors would like to acknowledge networking support by the COST Action CA18108
and funding support from Cosmology@MALTA which is supported by the University of Malta.

\begin{appendix}

\section{Derivation for the General Taylor Expansion model}
\label{app:GTE}
\noindent For this model, the Lagrangian turns out to be
\begin{equation}
f(T,B)\simeq A_0+A_1 T + A_2 T^2 + A_3 B^2 + A_4 TB\,,
\end{equation}
where if we use Eqs.(\ref{torsionscalar_frw})-(\ref{boundaryscalar_frw}), the EoS in Eq.(\ref{EoS_func}) can be related to the Hubble (or scale) factor directly. These derivatives are given by
\begin{eqnarray}
f_T &=& A_1 + 2A_2 T + A_4 B\,,\\
\dot{f}_T &=& 24 A_2 H\dot{H} + 6 A_4 \left(6H\dot{H} + \ddot{H}\right)\,,\\
f_B &=& 2A_3 B + A_4 T\,,\\
\dot{f}_B &=& 12 A_3 \left(6H\dot{H} + \ddot{H}\right) + 12 A_4 H\dot{H}\,,\\
\ddot{f}_B &=& 12A_3 \left(6\dot{H}^2 + 6 H \ddot{H} + \dddot{H}\right) + 12 A_4\left(\dot{H}^2 + H\ddot{H}\right)\,,
\end{eqnarray}
where the torsion scalar and boundary term represent the quantities in Eq.(\ref{torsionscalar_frw}) and Eq.(\ref{boundaryscalar_frw}) respectively. 

\section{Derivation for the Power Law model}
\label{app:PL}
\noindent For this model, the Lagrangian turns out to be
\begin{equation}
f(T,B) = b_0 B^k + t_0 T^m\,, 
\end{equation}
where the contributing elements to the EoS parameter in Eq.(\ref{EoS_func}) are given by
\begin{eqnarray}
f_T &=& t_0mT^{m-1}\,,\\
\dot{f}_T &=& 12t_0m(m-1)H\dot{H}\left(6H^2\right)^{m-2}\,,\\
f_B &=& b_0kB^{k-1}\,,\\
\dot{f}_B &=& 6b_0k(k-1)\left[6(3H^2+\dot{H})\right]^{k-2}\left(6H\dot{H}+\ddot{H}\right)\,,\\
\ddot{f}_B &=& 6b_0k(k-1)\left[6(3H^2+\dot{H})\right]^{k-2}\nonumber\\
&&\left\{\frac{k-1}{3H^2+\dot{H}}\left(6H\dot{H}+\ddot{H}\right)+6\dot{H}^2+6H\ddot{H}+\dddot{H}\right\}\,,
\end{eqnarray}
where it can be seen that the torsion scalar embodies the second-order elements of the $f(R)$ gravity while the boundary term $B$ takes on its fourth-order contributions.

\section{Derivation for the Mixed Power Law Model }
\label{app:MPL}
\noindent \noindent This models takes the form
\begin{equation}
f(T,B) = f_0 B^k T^m\,,
\end{equation}
where the second- and fourth-order contributions will now be mixed, and $f_0,k,m$ are arbitrary constants. In order to compare with the Friedmann equations, consider the derivatives below
\begin{eqnarray}
f_T &=& f_0 m B^k T^{m-1}\,,\\
\dot{f}_T &=& 6 f_0 m B^{k-1} T^{m-2} \left[kT\left(6H\dot{H} + \ddot{H}\right) + 2H\dot{H} (m-1) B\right]\,,\\
f_B &=& f_0 k B^{k-1} T^m\,,\\
\dot{f}_B &=& 6 f_0 k B^{k-2} T^{m-1} \left[(k-1)(6H\dot{H} + \ddot{H})T + 2mBH\dot{H}\right]\,,\\
\ddot{f}_B &=& 6f_0 k \Bigg\{ \left[(m-1) B^{k-2}T^{m-2} + 6(k-2)B^{k-3} T^{m-1} \left(6H\dot{H} + \ddot{H}\right)\right]\nonumber\\
&&\left[2mBH\dot{H} + (k-1)\left(6H\dot{H} + \ddot{H}\right)\right]\nonumber\\
&&+B^{k-2} T^{m-1} \Big[12H\dot{H}(k-1)\left(6H\dot{H} + \ddot{H}\right) + (k-1)T\left(6\dot{H}^2+6H\ddot{H} + \dddot{H}\right)\nonumber\\
&& + 2m\left(6H\dot{H}\left(6H\dot{H} + \ddot{H}\right) + B\dot{H}^2 + BH\ddot{H}\right)\Big]\Bigg\}\,,
\end{eqnarray}
where again the torsion scalar and boundary term represent the quantities in Eq.(\ref{torsionscalar_frw}) and Eq.(\ref{boundaryscalar_frw}) respectively.

\section{Boundary Term Deviations to TEGR model}
\label{app:TEGR}
\noindent In general, we can write the modifications to TEGR as
\begin{equation} 
f(T,B) = -T + g(B)\,,
\end{equation}
with $g(B) = f_1 B\ln B$ and where $f_1$ is an arbitrary constant. In this circumstance, the derivative quantities in the Friedmann equations are given by
\begin{eqnarray}
f_T &=& -1\,,\\
\dot{f}_T &=& 0\,,\\
f_B &=& f_1\ln B + f_1\,,\\
\dot{f}_B &=& 6f_1\,\frac{6H\dot{H} + \ddot{H}}{B}\,,\\
\ddot{f}_B &=& \frac{6f_1}{B^2}\,\left[\left(6\dot{H}^2 + 6 H\ddot{H} + \dddot{H}\right)\,B - 6\left(6H\dot{H} + \ddot{H}\right)^2\right]\,.
\end{eqnarray}

\end{appendix}
\bibliographystyle{JHEP}

\providecommand{\href}[2]{#2}\begingroup\raggedright\endgroup

\end{document}